\documentclass[english,aps,print,showpacs]{revtex4}
\usepackage[T1]{fontenc}
\usepackage[latin9]{inputenc}
\usepackage{graphicx}
\usepackage{amssymb}
\makeatletter

\providecommand{\LyX}{L\kern-.1667em\lower.25em\hbox{Y}\kern-.125emX\@}
\DeclareRobustCommand*{\lyxarrow}{%
\@ifstar
{\leavevmode\,$\triangleleft$\,\allowbreak}
{\leavevmode\,$\triangleright$\,\allowbreak}}

\@ifundefined{textcolor}{}
{%
 \definecolor{BLACK}{gray}{0}
 \definecolor{WHITE}{gray}{1}
 \definecolor{RED}{rgb}{1,0,0}
 \definecolor{GREEN}{rgb}{0,1,0}
 \definecolor{BLUE}{rgb}{0,0,1}
 \definecolor{CYAN}{cmyk}{1,0,0,0}
 \definecolor{MAGENTA}{cmyk}{0,1,0,0}
 \definecolor{YELLOW}{cmyk}{0,0,1,0}
}

\makeatother

\usepackage{babel}
\begin{document}

\title{Resonance fluorescence of strongly driven two-level system coupled to multiple dissipative baths}
\author{Yiying Yan\footnote{yiyingyan@sjtu.edu.cn}}
\author{Zhiguo L\"{u}\footnote{zglv@sjtu.edu.cn}}
\author{Hang Zheng\footnote{hzheng@sjtu.edu.cn}}
\affiliation{Key Laboratory of Artificial Structures and Quantum Control (Ministry of Education), Department of Physics and Astronomy,Shanghai Jiao Tong University, Shanghai 200240, China}
\affiliation{Collaborative Innovation Center of Advanced Microstructures, Nanjing 210093, China}
\date{\today}

\begin{abstract}
We present a theoretical formalism for resonance fluorescence radiating from a two-level system (TLS) driven by any periodic driving and coupled to multiple reservoirs. The formalism is derived analytically based on the combination of Floquet theory and Born-Markov master equation. The formalism allows us to calculate the spectrum when the Floquet states and quasienergies are analytically or numerically solved for simple or complicated driving fields. We can systematically explore the spectral features by implementing the present formalism. To exemplify this theory, we apply the unified formalism to comprehensively study a generic model that a harmonically driven TLS is simultaneously coupled to a radiative reservoir and a dephasing reservoir. We demonstrate that the significant features of the fluorescence spectra, the driving-induced asymmetry and the dephasing-induced asymmetry, can be attributed to the violation of detailed balance condition, and explained in terms of the driving-related transition quantities between Floquet-states and their steady populations. In addition, we find the distinguished features of the fluorescence spectra under the biharmonic and multiharmonic driving fields in contrast with that of the harmonic driving case. In the case of the biharmonic driving, we find that the spectra is significantly different from the result of the rotating-wave approximation (RWA) under the multiple resonance conditions. By the three concrete applications, we illustrate that the present formalism provides a routine tool for comprehensively exploring the fluorescence spectrum of periodically strongly driven TLSs.
\end{abstract}
\maketitle

\section{Introduction}
In recent years, the resonance fluorescence has attracted widespread attentions both in experiments and theory~\cite{mollow1969power,keitel1995resonance,wodkiewicz1976markovian,felinto2003single,muller2007resonance,ates2009post,flagg2009resonantly,vamivakas2009spin,ulrich2011dephasing,ulhaq2012cascaded,wrigge2008efficient,astafiev2010resonance,roy2011phonon,
moelbjerg2012resonance,ulhaq2013detuning,mccutcheon2013model,yan2013effects,konthasinghe2014resonant}, which is not only motivated by testing fundamental quantum optics theory
but also for the purposes of developing single quantum emitters for quantum light
spectroscopy and quantum information applications. In general, the spectrum of the fluorescence light is of primary interest, which can be calculated in theory and measured in experiments. It is well known that the fluorescence spectrum consists of coherent and incoherent components. The coherent one results from the elastic scattering while the incoherent one from the inelastic scattering~\cite{mollow1969power}. In particular, in the case of a two-level system (TLS) driven by the monochromatic driving, the incoherent part is made up of three split peaks, known as Mollow triplet. The generation of Mollow triplet can be understood physically by using the so-called dressed atom model~\cite{cohen1992atom}, which combines the TLS and the driving field. Within this model, the Mollow triplet is simply interpreted as a result of the transitions between the specific dressed states.

As resonance fluorescence in versatile experimental conditions are explored, the theory about radiation from the TLS has been intensively investigated and developed in two main ways. One is concerning with elaborated driving field other than the monochromatic one, for instance, optical pulse~\cite{moelbjerg2012resonance} and polychromatic driving field~\cite{konthasinghe2014resonant,ficek1993resonance,park2002dressed}, etc. In these cases, the problem becomes complicated due to the applied driving field. The other involves different kind of reservoirs and the simple monochromatic driving. It focuses the influence of the reservoirs coupled to the TLS on the spectral properties, such as dephasing coupling~\cite{ulhaq2013detuning,mccutcheon2013model} and the narrowband vacuum~\cite{keitel1996vacuum,keitel1995resonance}, etc.
To our knowledge, there is no general theoretical formalism to simultaneously reveal effects of both the arbitrary driving and the multiple reservoirs on the resonance fluorescence. In this work, we present a theoretical formalism that provides a unified treatment of resonance fluorescence spectrum of the TLS driven by an arbitrary periodic driving and coupled to multiple reservoirs.

As is well known, we can solve the time evolution of the periodically driven TLS by the Floquet theory in the absence of the reservoirs~\cite{shirley1965solution,sambe1973steady,grifoni1998driven}. On the other hand, provided that a TLS interacts weakly with the reservoirs,
we can apply the Born-Markov master-equation approach to get the reduced dynamics of the TLS~\cite{breuer2002theory}. In the presence of the periodic driving and reservoirs, it is feasible to combine the Floquet theory and the Born-Markov master equation into the Floquet-Born-Markov (FBM) master equation~\cite{grifoni1998driven,blumel1991dynamical,gramich2014lamb}. It is noticeable that, in Ref.~\cite{szczygielski2013markovian}, the authors show that the FBM master equation is consistent with the second law of thermodynamics under strong driving conditions in which case the traditional quantum optical master equation becomes inapplicable and is inconsistent with the second law~\cite{geva1995relaxation}. Intuitively, we can conclude that the FBM master equation provides the basis of unified treatment of fluorescence spectrum in the cases of periodic strong driving and multiple reservoirs.

In this paper, we present a formalism of fluorescence spectrum based on the FBM master equation, which is applicable to the situation where the TLS is periodically driven and weakly coupled to multiple reservoirs. When the Floquet state of the driven TLS is solved by the numerical technique or analytical method, the formalism allows us to not only straightforward calculate the spectrum but also explore the spectral features with a simple selection rule. In Sec.~\ref{appli1}, to exemplify the theory, we apply formal spectrum to the harmonically driven TLS, which weakly interacts with both the radiative and dephasing reservoirs. Based on the analytical results of fluorescence spectra, we demonstrate two kinds of asymmetric line shape in the main triplet. One is the driving-induced asymmetric line shape, the other is the dephasing-induced asymmetric line shape. Moreover, we explain the underlying reason of the asymmetric lineshape and observe the interplay between the two kinds of asymmetry. In Sec.~\ref{appli2}, we study the fluorescence spectra of the TLS driven by a biharmonic driving. In comparison with the harmonic driving, the biharmonic driving leads to more intense higher-order triplets centered at both even and odd multiples of the driving frequency. We analyze the feature of the multiple resonance induced fluorescence spectra. In such case, we find the exotic spectra significantly differ from those predicted by the rotating-wave approximation (RWA). In Sec.~\ref{appli3}, we discuss briefly the fluorescence spectra of the multiharmonic driving similar to the square-wave signal. It turns out that the present formalism provides a routine tool for the comprehensive studies of the fluorescence spectrum.

\section{Theoretical formalism}\label{sec:general}
\subsection{Floquet-Born-Markov master equation\label{sub:FBM}}

We consider a TLS with time-dependent periodic
driving and multiple reservoirs (including a radiation field), which is described
by the Hamiltonian (we set $\hbar=1$ throughout this paper)
\begin{equation}
H(t)=H_{S}(t)+\sum_{j}H_{R}^{(j)}+\sum_{j}H_{SR}^{(j)}.\label{eq:H0}
\end{equation}
Here, $H_{S}(t)=H_{S}(t+2\pi/\omega_{l})$ is the periodically
driven TLS. $\sum_{j}H_{R}^{(j)}\equiv H_{R}$
denotes the sum of free Hamiltonians of the reservoirs. $\sum_{j}H_{SR}^{(j)}\equiv H_{SR}$ represents the coupling
between the TLS and reservoirs. Moreover, we assume that $H_{SR}^{(j)}\equiv\hat{X}^{(j)}\otimes\hat{B}^{(j)}$,
where $\hat{X}^{(j)}$ and $\hat{B}^{(j)}$ are the Hermite operators
and act on the Hilbert space of the TLS and that of the $j$th reservoir, respectively.

Since $H_{S}(t)$ is periodic in time, we can use the Floquet theorem
to solve the dynamics of the TLS. The Theorem states that the Schr\"{o}dinger
equation governed by $H_{S}(t)$ possesses the formal solution~\cite{shirley1965solution,sambe1973steady,grifoni1998driven}:
\begin{equation}
|\psi_{\alpha}(t)\rangle=e^{-i\varepsilon_{\alpha}t}|u_{\alpha}(t)\rangle,\label{eq:formalsol}
\end{equation}
where $|u_{\alpha}(t)\rangle=|u_{\alpha}(t+2\pi/\omega_{l})\rangle$, a function
periodic in time, is referred to as the Floquet state associated with
quasienergy $\varepsilon_{\alpha}$. It is straightforward to show
that $|u_{\alpha}(t)\rangle$ and $\varepsilon_{\alpha}$ satisfy
following equation:
\begin{equation}
[H_{S}(t)-i\partial_{t}]|u_{\alpha}(t)\rangle=\varepsilon_{\alpha}|u_{\alpha}(t)\rangle,\label{eq:eigeneq}
\end{equation}
where $H_{S}(t)-i\partial_{t}$ is the so-called Floquet Hamiltonian.
It is worthwhile to notice that $|u_{\alpha n}(t)\rangle=e^{in\omega_{l}t}|u_{\alpha}(t)\rangle$
is physically equivalent to $|u_{\alpha}(t)\rangle$
but with the shifted quasienergy $\varepsilon_{\alpha n}=\varepsilon_{\alpha}+n\omega_{l}$.
As a consequence, it is sufficient to consider $\varepsilon_{\alpha}$
in the range $-\frac{\omega_{l}}{2}<\varepsilon_{\alpha}\leq\frac{\omega_{l}}{2}$.
In what follows, we revisit the derivation of Born-Markov master in
the Floquet picture.

In the weak-coupling regime, we set $H_{0}=H_{S}(t)+H_{R}$ as
the free Hamiltonian and $H_{SR}$ as the perturbation. In the interaction
picture, we readily obtain the Born-Markov master equation for the
TLS up to second order of the perturbation~\cite{breuer2002theory}, which reads
\begin{equation}
\frac{d}{dt}\rho_{S}^{\mathrm{I}}(t)=-\int_{0}^{\infty}d\tau\mathrm{Tr}_{R}[H_{\mathrm{I}}(t),[H_{\mathrm{I}}(t-\tau),\rho_{S}^{\mathrm{I}}(t)\rho_{R}]],\label{eq:meo}
\end{equation}
Here, $\rho_{S}^{\mathrm{I}}(t)$ is the reduced density matrix of
the TLS in the interaction picture. $H_{{\rm I}}(t)$ is given by
\begin{eqnarray}
H_{{\rm I}}(t) & = & U_{S}^{\dagger}(t)\exp(iH_{R}t)H_{SR}U_{S}(t)\exp(-iH_{R}t)\nonumber \\
 & = & \sum_{j}U_{S}^{\dagger}(t)\hat{X}^{(j)}U_{S}(t)\otimes\exp(iH_{R}t)\hat{B}^{(j)}\exp(-iH_{R}t)\nonumber \\
 & \equiv & \sum_{j}\hat{X}^{(j)}(t)\otimes\hat{B}^{(j)}(t),
\end{eqnarray}
where $U_{S}(t)=\mathcal{T}\exp[-i\int_{0}^{t}H_{S}(\tau)d\tau]$
is the time-ordered evolution operator of the TLS. Provided that $[H_{R},\rho_{R}]=0$,
we can rewrite Eq.~(\ref{eq:meo}) as
\begin{eqnarray}
\frac{d}{dt}\rho_{S}^{\mathrm{I}}(t) & = & -\sum_{j}\int_{0}^{\infty}d\tau[\hat{X}^{(j)}(t)\hat{X}^{(j)}(t-\tau)\rho_{S}^{\mathrm{I}}(t)\langle \hat{B}^{(j)}(\tau)\hat{B}^{(j)}(0)\rangle _{R}\nonumber \\
 &  & -\hat{X}^{(j)}(t-\tau)\rho_{S}^{\mathrm{I}}(t)\hat{X}^{(j)}(t)\langle \hat{B}^{(j)}(\tau)\hat{B}^{(j)}(0)\rangle _{R}+\mathrm{h.c.}],\label{eq:fbmme}
\end{eqnarray}
where $\langle \hat{B}^{(j)}(\tau)\hat{B}^{(j)}(0)\rangle _{R}\equiv\mathrm{Tr}_{R}[\hat{B}^{(j)}(\tau)\hat{B}^{(j)}(0)\rho_{R}]$
is the reservoir correlation function. To proceed, we use the Floquet
states $|u_{\alpha}(0)\rangle$ ($\alpha=\pm$) as the basis to derive
the equation of motion for the element $\rho_{\alpha\beta}^{\mathrm{I}}(t)=\langle u_{\alpha}(0)|\rho_{S}^{\mathrm{I}}(t)|u_{\beta}(0)\rangle$.
According to Floquet theory, we have
\begin{equation}
\langle u_{\alpha}(0)|\hat{X}^{(j)}(t)|u_{\beta}(0)\rangle=\langle u_{\alpha}(t)|\hat{X}^{(j)}|u_{\beta}(t)\rangle e^{i(\varepsilon_{\alpha}-\varepsilon_{\beta})t}=\sum_{n}X_{\alpha\beta,n}^{(j)}e^{i\Delta_{\alpha\beta,n}t},
\end{equation}
where
\begin{eqnarray}
X_{\alpha\beta,n}^{(j)} & = & \frac{\omega_{l}}{2\pi}\int_{0}^{2\pi/\omega_{l}} dt  \langle u_{\alpha}(t)|\hat{X}^{(j)}|u_{\beta}(t)\rangle e^{-in\omega_{l}t} ,\label{eq:xabn}\\
\Delta_{\alpha\beta,n} & = & \varepsilon_{\alpha}-\varepsilon_{\beta}+n\omega_{l}.
\end{eqnarray}
Thus, we readily obtain following expressions:
\begin{eqnarray}
\langle u_{\alpha}(0)|\hat{X}^{(j)}(t)\hat{X}^{(j)}(t-\tau)\rho_{S}^{\mathrm{I}}(t)|u_{\beta}(0)\rangle & = & \sum_{\gamma,\delta}\langle u_{\alpha}(0)|\hat{X}^{(j)}(t)|u_{\gamma}(0)\rangle \langle u_{\gamma}(0)|\hat{X}^{(j)}(t-\tau)|u_{\delta}(0)\rangle\nonumber\\
& &\times \langle u_{\delta}(0)|\rho_{S}^{\mathrm{I}}(t)|u_{\beta}(0)\rangle\nonumber \\
 & = & \sum_{\gamma,\delta,n,m}\sum_{\lambda}\delta_{\beta,\gamma}X_{\alpha\lambda,n}^{(j)}X_{\lambda\delta,m}^{(j)}e^{-i\Delta_{\lambda\delta,m}\tau}\nonumber\\
& &\times\rho_{\delta\gamma}^{\mathrm{I}}(t)e^{i(\Delta_{\alpha\delta,n}+\Delta_{\gamma\beta,m})t},\label{eq:uab1}\\
\langle u_{\alpha}(0)|\hat{X}^{(j)}(t-\tau)\rho_{S}^{\mathrm{I}}(t)\hat{X}^{(j)}(t)|u_{\beta}(0)\rangle & = & \sum_{\gamma,\delta,n,m}X_{\alpha\delta,n}^{(j)}X_{\gamma\beta,m}^{(j)}e^{-i\Delta_{\alpha\delta,n}\tau}\nonumber\\
& &\times\rho_{\delta\gamma}^{\mathrm{I}}(t)e^{i(\Delta_{\alpha\delta,n}+\Delta_{\gamma\beta,m})t}.\label{eq:uab2}
\end{eqnarray}
Substituting Eqs.~(\ref{eq:uab1}) and (\ref{eq:uab2}) into Eq.~(\ref{eq:fbmme}),
we arrive at the following form
\begin{eqnarray}
\frac{d}{dt}\rho_{\alpha\beta}^{\mathrm{I}}(t) & = & \sum_{\delta,\gamma,n,m}\bigg\{\Gamma_{\gamma\beta\alpha\delta,mn}^{+}+\Gamma_{\gamma\beta\alpha\delta,mn}^{-}-\sum_{\lambda}\bigg[\delta_{\beta,\gamma}\Gamma_{\alpha\lambda\lambda\delta,nm}^{+}\nonumber \\
 &  & +\delta_{\alpha,\delta}\Gamma_{\gamma\lambda\lambda\beta,nm}^{-}\bigg]\bigg\}\rho_{\delta\gamma}^{\mathrm{I}}(t)e^{i(\Delta_{\gamma\beta,m}+\Delta_{\alpha\delta,n})t},\label{eq:fbm}
\end{eqnarray}
where
\begin{eqnarray}
\Gamma_{\alpha\beta\delta\gamma,nm}^{+} & = & \sum_{j}X_{\alpha\beta,n}^{(j)}X_{\delta\gamma,m}^{(j)}\gamma_{\delta\gamma,m}^{(j)+},\\
\Gamma_{\alpha\beta\delta\gamma,nm}^{-} & = & \sum_{j}X_{\alpha\beta,n}^{(j)}X_{\delta\gamma,m}^{(j)}\gamma_{\alpha\beta,n}^{(j)-},\\
\gamma_{\alpha\beta,n}^{(j)\pm} & = & \int_{0}^{\infty} d\tau e^{-i\Delta_{\alpha\beta,n}\tau}\langle\hat{B}^{(j)}(\pm\tau)\hat{B}^{(j)}\rangle_{R}.
\end{eqnarray}
This is the so-called Floquet-Born-Markov master equation~\cite{grifoni}. This master
equation can treat the driving term exactly and keep TLS-reservoir couplings up to
second order.

In some senses, i.e., the strong driving cases, we may neglect the
time-dependent terms in Eq.~(\ref{eq:fbm}) by invoking the secular
approximation. With such approximation~\cite{grifoni1998driven}, we obtain a
time-independent equation of motion in the Floquet picture,
\begin{equation}
\frac{d}{dt}\rho_{\alpha\beta}(t)=-i(\varepsilon_{\alpha}-\varepsilon_{\beta})\rho_{\alpha\beta}(t)-\gamma_{\alpha\beta}\rho_{\alpha\beta}(t)+\delta_{\alpha,\beta}\sum_{\delta\neq\beta}\rho_{\delta\delta}(t)W_{\delta\beta},\label{eq:secufbm}
\end{equation}
where we used relation $\rho_{\alpha\beta}(t)\equiv\langle u_{\alpha}(t)|\rho_{S}(t)|u_{\beta}(t)\rangle=e^{-i(\varepsilon_{\alpha}-\varepsilon_{\beta})t}\rho_{\alpha\beta}^{\mathrm{I}}(t).$
The coefficients read
\begin{eqnarray}
W_{\delta\beta} & = & \sum_{n,m}(\Gamma_{\delta\beta\beta\delta,mn}^{+}+\Gamma_{\delta\beta\beta\delta,mn}^{-})\delta_{n,-m},\\
\gamma_{\alpha\beta} & = & \sum_{n,m}\left[\sum_{\lambda}\left(\Gamma_{\alpha\lambda\lambda\alpha,nm}^{+}+\Gamma_{\beta\lambda\lambda\beta,nm}^{-}\right)-\Gamma_{\beta\beta\alpha\alpha,mn}^{+}-\Gamma_{\beta\beta\alpha\alpha,mn}^{-}\right]\delta_{n,-m},
\end{eqnarray}
where $\delta_{n,-m}$ is the Kronecker-delta function. The solutions to Eq.~(\ref{eq:secufbm}) can be found as follows:
\begin{eqnarray}
\rho_{++}(t) & = & \rho_{++}(0)e^{-\gamma_{{\rm rel}}t}+\frac{W_{-+}}{\gamma_{{\rm rel}}}(1-e^{-\gamma_{{\rm rel}}t}),\label{eq:sol1-1}\\
\rho_{+-}(t) & = & \rho_{+-}(0)e^{-(\gamma_{{\rm deph}}+i\varepsilon_{+}-i\varepsilon_{-}+i\delta\omega_{+-})t},\label{eq:sol2-1}
\end{eqnarray}
where the explicit forms of the transition rates and Lamb shift can be
rewritten as follows:
\begin{eqnarray}
W_{-+} & = & \sum_{n,j}|X_{+-,n}^{(j)}|^{2}(\gamma_{+-,n}^{(j)+}+\gamma_{-+,-n}^{(j)-}),\label{eq:transrate}\\
\gamma_{{\rm rel}} & = & \sum_{n,j}|X_{+-,n}^{(j)}|^{2}(\gamma_{+-,n}^{(j)+}+\gamma_{-+,-n}^{(j)-}+\gamma_{-+,-n}^{(j)+}+\gamma_{+-,n}^{(j)-}),\\
\gamma_{{\rm deph}} & = & \sum_{n,j}\big\{|X_{-+,n}^{(j)}|^{2}\mathrm{Re}(\gamma_{-+,n}^{(j)+}+\gamma_{-+,n}^{(j)-})\nonumber \\
 &  & +|X_{++,n}^{(j)}|^{2}(\gamma_{++,n}^{(j)+}+\gamma_{++,-n}^{(j)+}+\gamma_{--,n}^{(j)-}+\gamma_{--,-n}^{(j)-})\big\},\label{eq:dephrate}\\
\delta\omega_{+-} & = & \sum_{n,j}|X_{-+,n}^{(j)}|^{2}\mathrm{Im}(\gamma_{-+,n}^{(j)+}+\gamma_{-+,n}^{(j)-}).\label{eq:lambs-1}
\end{eqnarray}
Here $\gamma_{{\rm rel}}$ and $\gamma_{{\rm deph}}$ are the relaxation
and dephasing rates of the Floquet states, respectively. $\delta\omega_{+-}$
is the reservoirs-induced energy shift, which is usually a negligible
small quantity and omitted. In the next section, we use these solutions
to derive the fluorescence spectrum.

\subsection{Resonance fluorescence spectrum}

In this subsection we derive an analytical expression for the fluorescence spectrum
in the steady-state limit. The fluorescence spectrum is proportional to the real part of Fourier transform of the first-order correlation function~\cite{mollow1969power}
\begin{equation}
I(\omega)\propto\mathrm{Re}\int_{0}^{\infty}\lim_{t\rightarrow\infty}g^{(1)}(t+\tau,t)e^{-i\omega\tau}d\tau,\label{eq:definition}
\end{equation}
where $g^{(1)}(t+\tau,t)$ is the first-order correlation function
and evaluated as
\begin{eqnarray}
g^{(1)}(t+\tau,t) & = & \mathrm{Tr}[U^{\dagger}(t+\tau)\sigma_{+}U(t+\tau)U^{\dagger}(t)\sigma_{-}U(t)\rho_{S}(0)\rho_{R}]\nonumber \\
 & = & \mathrm{Tr}_{S}\{\sigma_{+}\mathrm{Tr}_{R}[U(t+\tau)U^{\dagger}(t)\sigma_{-}\rho_{S}(t)\rho_{R}U(t)U^{\dagger}(t+\tau)]\}\nonumber \\
 & \equiv & \mathrm{Tr}_{S}[\sigma_{+}\rho_{S}(t+\tau)],
\end{eqnarray}
where $\rho_{S}(t+\tau)=\mathrm{Tr}_{R}[U(t+\tau)U^{\dagger}(t)\sigma_{-}\rho_{S}(t)\rho_{R}U(t)U^{\dagger}(t+\tau)]$
can be viewed as a reduced density matrix whose evolution is determined
by Eq.~(\ref{eq:secufbm}) with initial condition $\sigma_{-}\rho_{S}(t)$.
This is actually the so-called quantum regression theorem~\cite{lax1968quantum}. In general,
we have
\begin{equation}
\sigma_{-}\rho_{S}(t)=\sum_{\alpha,\gamma,\beta,n}X_{\alpha\gamma,n}^{-}\rho_{\gamma\beta}(t)|u_{\alpha}(t)\rangle\langle u_{\beta}(t)|e^{in\omega_{l}t},
\end{equation}
where
\begin{equation}
  X^-_{\alpha\beta,n}=\frac{\omega_l}{2\pi}\int^{2\pi/\omega_l}_0 dt\langle u_\alpha(t)|\sigma_-|u_{\beta}(t)\rangle e^{-in\omega_l t}.
\end{equation}
According to quantum regression theorem~\cite{lax1968quantum},
the explicit form of $\rho_{S}(t+\tau)$ can be obtained from Eqs.~(\ref{eq:sol1-1}) and (\ref{eq:sol2-1}) by replacing the initial condition $\rho_{\alpha\beta}(0)\rightarrow\sum_{\gamma,n}X_{\alpha\gamma,n}^{-}\rho_{\gamma\beta}(t)e^{in\omega_{l}t}$,
which leads to
\begin{eqnarray}
\rho_{++}(t+\tau) & = & \sum_{\gamma,n}X_{+\gamma,n}^{-}\rho_{\gamma+}(t)e^{in\omega_{l}t}e^{-\gamma_{{\rm rel}}\tau}+\rho_{++}^{{\rm ss}}(1-e^{-\gamma_{{\rm rel}}\tau})\sum_{\alpha,\gamma,n}X_{\alpha\gamma,n}^{-}\rho_{\gamma\alpha}(t)e^{in\omega_{l}t},\\
\rho_{--}(t+\tau) & = & \sum_{\gamma,n}X_{-\gamma,n}^{-}\rho_{\gamma-}(t)e^{in\omega_{l}t}e^{-\gamma_{{\rm rel}}\tau}+\rho_{--}^{{\rm ss}}(1-e^{-\gamma_{{\rm rel}}\tau})\sum_{\alpha,\gamma,n}X_{\alpha\gamma,n}^{-}\rho_{\gamma\alpha}(t)e^{in\omega_{l}t},\\
\rho_{+-}(t+\tau) & = & \sum_{\gamma,n}X_{+\gamma,n}^{-}\rho_{\gamma-}(t)e^{in\omega_{l}t}e^{-(\gamma_{{\rm deph}}+i\omega_{+-})\tau},\\
\rho_{-+}(t+\tau) & = & \sum_{\gamma,n}X_{-\gamma,n}^{-}\rho_{\gamma+}(t)e^{in\omega_{l}t}e^{-(\gamma_{{\rm deph}}-i\omega_{+-})\tau},
\end{eqnarray}
where $\omega_{+-}=\varepsilon_{+}-\varepsilon_{-}+\delta\omega_{+-}$
are the reservoirs-normalized energy gap of the Floquet states. $\rho_{++}^{{\rm ss}}=\frac{W_{-+}}{\gamma_{{\rm rel}}}$
and $\rho_{--}^{{\rm ss}}=1-\rho_{++}^{{\rm ss}}$ are the steady
Floquet-state populations. In the steady-state limit,
we can assume the correlation function to be $\tau$ dependent only,
and thus it can be rewritten as
\begin{eqnarray}
g^{(1)}(\tau) & = & \lim_{t\rightarrow\infty}g^{(1)}(t+\tau,t)\nonumber \\
 & = & \lim_{t\rightarrow\infty}\sum_{\alpha,\beta,n}X_{\alpha\beta,n}^{+}\rho_{\beta\alpha}(t+\tau)e^{in\omega_{l}(t+\tau)}\nonumber \\
 & = & \sum_{n}e^{in\omega_{l}\tau}|X_{++,n}^{+}|^{2}\{[1-(\rho_{++}^{{\rm ss}}-\rho_{--}^{{\rm ss}})^{2}]e^{-\gamma_{{\rm rel}}\tau}+(\rho_{++}^{{\rm ss}}-\rho_{--}^{{\rm ss}})^{2}\}\nonumber \\
 &  & +\sum_{n}e^{in\omega_{l}\tau}\{|X_{-+,n}^{+}|^{2}\rho_{--}^{{\rm ss}}e^{-(\gamma_{{\rm deph}}+i\omega_{+-})\tau}+|X_{+-,n}^{+}|^{2}\rho_{++}^{{\rm ss}}e^{-(\gamma_{{\rm deph}}-i\omega_{+-})\tau}\},\label{eq:gtau}
\end{eqnarray}
where $X_{\alpha\beta,n}^{+}=(X_{\beta\alpha,-n}^{-})^{\ast}.$ By
substituting Eq.~(\ref{eq:gtau}) into Eq.~(\ref{eq:definition})
and integrating, we obtain the fluorescence spectrum as follows:
\begin{eqnarray}
I(\omega) & \propto & \sum_{n}\bigg\{\pi|X_{++,n}^{+}|^{2}(\rho_{++}^{{\rm ss}}-\rho_{--}^{{\rm ss}})^{2}\delta(\omega-n\omega_{l})\nonumber \\
 &  & +|X_{++,n}^{+}|^{2}[1-(\rho_{++}^{{\rm ss}}-\rho_{--}^{{\rm ss}})^{2}]\frac{\gamma_{{\rm rel}}}{\gamma_{{\rm rel}}^{2}+(\omega-n\omega_{l})^{2}}\nonumber \\
 &  & +|X_{-+,n}^{+}|^{2}\rho_{--}^{{\rm ss}}\frac{\gamma_{{\rm deph}}}{\gamma_{{\rm deph}}^{2}+(\omega-n\omega_{l}+\omega_{+-})^{2}}\nonumber \\
 &  & +|X_{+-,n}^{+}|^{2}\rho_{++}^{{\rm ss}}\frac{\gamma_{{\rm deph}}}{\gamma_{{\rm deph}}^{2}+(\omega-n\omega_{l}-\omega_{+-})^{2}}\bigg\}.\label{eq:sfun}
\end{eqnarray}
This expression provides the unified description of fluorescence in the cases of the periodically driven TLS coupled to multiple reservoirs.
We state that this result is derived with the aid of
the secular approximation, and thus it is valid when $|\omega_{+-}|\gg\gamma_{\rm deph}, \gamma_{\rm rel}$, i.e., the emission lines are well separated. While this condition is not satisfied, we can calculate the fluorescence spectrum from Eq.~(\ref{eq:fbm}) by retaining the terms satisfying $n=-m$, i.e., the partial secular approximation.

In the present formalism, we can interpret the emission lines as a
result of the transitions of specific Floquet states and understand the physical origin of the emission
line. Generally speaking, a transition
$|u_{\alpha,n}(t)\rangle \rightarrow |u_\beta(t)\rangle$ results in
an emission process with the resulting photon of frequency determined
by the energy gap of the two states (a positive frequency means a real process)
and probability related to $|X^+_{\alpha\beta,n}|^2$.
The $X^+_{\alpha\beta,n}$ is defined as
\[
 X^+_{\alpha\beta,n}=(X^-_{\beta\alpha,-n})^\ast=\frac{\omega_l}{2\pi}\int^{2\pi/\omega_l}_{0}dt\langle u_{\alpha,n}(t)|\sigma_+|u_\beta(t)\rangle,
\]
which can be regarded as the time-averaged transition amplitude between the two states,
$|u_{\alpha n}(t)\rangle$ and $|u_{\beta}(t)\rangle$. Moreover, it is
straightforward to calculate the quasienergy gap between the two
states
\[
\Delta_{\alpha\beta,n}=\varepsilon_{\alpha,n}-\varepsilon_{\beta}=\omega_{\alpha\beta}+n\omega_l,
\]
where
$\omega_{\alpha\beta}\equiv\varepsilon_\alpha-\varepsilon_\beta$. All results indicate that the weights of Lorentzian lines in the formal spectrum depend on $|X^+_{\alpha\beta,n}|^2$ and the populations of the Floquet states, and the positions are determined by $\Delta_{\alpha\beta,n}$. For instance, the incoherent Lorentzian
line
\[
\frac{\gamma_{{\rm deph}}}{\gamma_{{\rm
deph}}^{2}+(\omega-n\omega_{l}+\omega_{+-})^{2}}
\]
is corresponding to the transition $|u_{-,n}(t)\rangle
\rightarrow |u_{+}(t)\rangle$. Therefore, its weight is
proportional to the transition probability $|X^{+}_{-+,n}|^2$ as
well as the population of the Floquet state $\rho^{\rm{ss}}_{--}$.
Its position is determined by the gap
$\varepsilon_{-,n}-\varepsilon_{+}=\omega_{-+}+n\omega_l\equiv
n\omega_l-\omega_{+-}$. $\gamma_{\rm{deph}}$ is the dephasing rate
for the Floquet state, which determines the FWHM. The other emission
lines can be understood in the same manner.

It turns out that the semiclassical Floquet theory provides a similar description as the quantum dressed-atom model. In fact, it has been discussed in Ref.~\cite{breuer1997dissipative} that the correspondence between
the Floquet states and the dressed states for the simple RWA driving case. All in all, the present formalism allows us to obtain analytical expressions for all the characteristics (weights and widths) of the fluorescence spectrum and physically understand how the driving and reservoirs influence spectral characteristics.
In what follows, we use three examples to show the advantages of the present formalism.

\section{Application to the harmonic driving}\label{appli1}

We study the fluorescence of a specific model by implementing the formalism presented above. The model describes that a TLS is excited by a monochromatic harmonic field and weakly coupled to an electromagnetic radiation field and a dephasing reservoir. The total Hamiltonian reads
\begin{eqnarray}
H(t) & = & H_{S}(t)+\sum_{\mathbf{k}}\omega_{\mathbf{k}}a_{\mathbf{k}}^{\dagger}a_{\mathbf{k}}+\sum_{\mathbf{q}}\nu_{\mathbf{q}}b_{\mathbf{q}}^{\dagger}b_{\mathbf{q}}\nonumber \\
 &  & +\frac{\sigma_{x}}{2}\sum_{\mathbf{k}}g_{\mathbf{k}}(a_{\mathbf{k}}^{\dagger}+a_{\mathbf{k}})+\frac{\sigma_{z}}{2}\sum_{\mathbf{q}}f_{\mathbf{q}}(b_{\mathbf{q}}^{\dagger}+b_{\mathbf{q}}).\label{eq:model}
\end{eqnarray}
Here, $H_{S}(t)=\frac{1}{2}\omega_{0}\sigma_{z}+\frac{A}{2}\cos(\omega_{l}t)\sigma_{x}$
describes that a TLS with bare transition frequency $\omega_{0}$
is driven by the harmonic field with amplitude $A$ and frequency $\omega_{l}$.
$\sigma_{x(y,z)}$ denotes Pauli matrix. $a_{{\bf k}}$ ($a_{{\bf k}}^{\dagger}$)
is the annihilation (creation) operator of the ${\bf k}$th-mode of electromagnetic
field with photon frequency $\omega_{{\bf k}}$. $b_{{\bf q}}$ ($b_{{\bf q}}^{\dagger}$)
is the annihilation (creation) operator of the ${\bf k}$th-mode with
frequency $\nu_{{\bf q}}$ of the dephasing reservoir. $g_{{\bf k}}$
and $f_{{\bf q}}$ are the coupling constants between TLS and reservoirs.

We state that the key ingredients to be determined in the present formalism are the Floquet states, quasienergies,
the coefficient $X_{\alpha\beta,n}^{(j)}$, and the Fourier transform
of the reservoir correlation function $\gamma_{\alpha\beta,n}^{(j)\pm}$.
These quantities fully determine the dynamics of the driven TLS as
well as the fluorescence spectrum given in Eq.~(\ref{eq:sfun}). For the model considered, in comparison with Eq.~(\ref{eq:H0}), we have $H_{R}=\sum_{{\bf k}}\omega_{{\bf k}}a_{{\bf k}}^{\dagger}a_{{\bf k}}+\sum_{{\bf q}}\nu_{{\bf q}}b_{{\bf q}}^{\dagger}b_{{\bf q}}$
and $H_{SR}=\sum_{j=x,z}\hat{X}^{(j)}\otimes\hat{B}^{(j)}$, where $\hat{X}^{(j)}=\sigma_{j}/2$,
$\hat{B}^{x}=\sum_{{\bf k}}g_{{\bf k}}(a_{{\bf k}}^{\dagger}+a_{{\bf k}})$,
and $\hat{B}^{z}=\sum_{{\bf q}}f_{{\bf q}}(b_{{\bf q}}^{\dagger}+b_{{\bf q}}).$
Provided that the Floquet states and quasienergies are known, we can determine the quantities listed in Eqs.~(\ref{eq:transrate})-(\ref{eq:lambs-1}) by the explicit
forms of the operators. Therefore, the key task is to determine the Floquet states and quasienergies. In
what follows, we use a unitary transformation to solve the Floquet
states and quansienergies for the harmonically driven TLS.

\subsection{Unitary transformation}

We start to derive the Floquet states and quasienergies of the harmonically driven TLS. This issue has been addressed by Shirley using the perturbation
treatment in $A$~\cite{shirley1965solution}. Here, we introduce a method based on a unitary transformation to analytically solve the Floquet states and quasienergies. This is the advantage of our treatment avoiding the diagonalization of the infinite Floquet Hamiltonian given by Shirley\cite{lu2012effects,shirley1965solution}.

The time evolution of any driven systems satisfies the following equation:
\begin{equation}
i\frac{d}{dt}U_{S}(t)=H_{S}(t)U_{S}(t),
\end{equation}
with the initial condition $U_{S}(0)=1$. By unitary transformation, we get
\begin{equation}
i\frac{d}{dt}U_{S}^{\prime}(t)=H_{S}^{\prime}(t)U_{S}^{\prime}(t),
\end{equation}
where $U_{S}^{\prime}(t)=e^{S(t)}U_{S}(t)$ and
\begin{equation}
H_{S}^{\prime}(t)=e^{S(t)}H_{S}(t)e^{-S(t)}+i\partial_{t}S(t)
\end{equation}
is the transformed Hamiltonian. The generator $S(t)$ of our treatment is
\begin{equation}
S(t)=i\frac{A}{2\omega_{l}}\xi\sin(\omega_{l}t)\sigma_{x},
\end{equation}
where the parameter $\xi$ can be determined
self-consistently ($0 \leq \xi\leq 1$)~\cite{lu2012effects}. We readily give the transformed Hamiltonian as
follows:
\begin{eqnarray}
H^{\prime}_{S}(t) & = & \frac{1}{2}\omega_{0}\left\{ \cos\left[\frac{A\xi}{\omega_{l}}\sin(\omega_{l}t)\right]\sigma_{z}+\sin\left[\frac{A\xi}{\omega_{l}}\sin(\omega_{l}t)\right]\sigma_{y}\right\} \nonumber \\
 &  & +\frac{A}{2}(1-\xi)\cos(\omega_{l}t)\sigma_{x}.
\end{eqnarray}
Using the identity $\exp\left[i\frac{A\xi}{\omega}\sin(\omega t)\right]=\sum_{n=-\infty}^{\infty}J_{n}\left(\frac{A\xi}{\omega}\right)\exp(in\omega t),$
in which $J_{n}(\cdot)$ is the $n$th-order Bessel function of the
first kind, we divide the transformed Hamiltonian into two parts
$H_{S}^{\prime}(t)=H_{{1}}^{\prime}(t)+H_{2}^{\prime}(t)$,
\begin{eqnarray}
H_{{1}}^{\prime}(t) & = & \frac{1}{2}\omega_{0}J_{0}\left(\frac{A\xi}{\omega_{l}}\right)\sigma_{z}+\omega_{0}J_{1}\left(\frac{A\xi}{\omega_{l}}\right)\sin(\omega_{l}t)\sigma_{y}+\frac{A}{2}(1-\xi)\cos(\omega_{l}t)\sigma_{x},\\
H_{2}^{\prime}(t) & = & \omega_{0}\sum_{n=1}^{\infty}\left\{ J_{2n}\left(\frac{A\xi}{\omega_{l}}\right)\cos(2n\omega_{l}t)\sigma_{z}+J_{2n+1}\left(\frac{A\xi}{\omega_{l}}\right)\sin[(2n+1)\omega_{l}t]\sigma_{y}\right\} .
\end{eqnarray}
We emphasis that $H_{{1}}^{\prime}(t)$ is comprised of the
slow-oscillating terms while $H_{2}^{\prime}(t)$ consists of all the
fast-oscillating terms. We introduce the first approximation in our
treatment, i.e., the drop of $H_{2}^{\prime}(t)$
and the Hamiltonian $H_{S}^{\prime}(t)\simeq H_{{1}}^{\prime}(t)$.
To proceed, we determine $\xi$ self-consistently by
\begin{equation}
\omega_{0}J_{1}\left(\frac{A\xi}{\omega_{l}}\right)=\frac{A}{2}(1-\xi)\equiv\frac{\tilde{A}}{4},
\end{equation}
which leads to a counter-rotating hybridized rotating-wave (CHRW)
Hamiltonian $H_{{\rm CHRW}}^{\prime}(t)$:
\begin{equation}
H_{{\rm CHRW}}^{\prime}(t)=\frac{1}{2}J_{0}\left(\frac{A\xi}{\omega_{l}}\right)\omega_{0}\sigma_{z}+\frac{\tilde{A}}{4}(e^{-i\omega_{l}t}\sigma_{+}+e^{i\omega_{l}t}\sigma_{-}),
\end{equation}
where $\sigma_{\pm}=(\sigma_{x}\pm i\sigma_{y})/2.$

The effective Hamiltonian $H_{{\rm CHRW}}^{\prime}(t)$ can be further
transformed into a time-independent form with a rotating operation
\begin{equation}
H_{{\rm CHRW}}^{\prime}=R(t)H_{{\rm CHRW}}^{\prime}(t)R^{\dagger}(t)-iR(t)\partial_{t}R^{\dagger}(t)=\frac{\tilde{\Delta}}{2}\sigma_{z}+\frac{\tilde{A}}{4}\sigma_{x},\label{eq:Happ}
\end{equation}
where $R(t)=\exp(i\sigma_{z}\omega_{l}t/2)$ and $\tilde{\Delta}=J_{0}\left(\frac{A\xi}{\omega_{l}}\right)\omega_{0}-\omega_{l}$
is the effective detuning. We can readily diagonalize the Hamiltonian
(\ref{eq:Happ}). Its eigenstates and corresponding eigenenergies
are given as follows:
\begin{eqnarray}
|\widetilde{\pm}\rangle & = & \sin\theta|\mp\rangle\pm\cos\theta|\pm\rangle,\\
E_{\pm} & = & \pm\frac{1}{2}\sqrt{\tilde{\Delta}^{2}+\tilde{A}^{2}/4}\equiv\pm\frac{1}{2}\tilde{\Omega}_{R},
\end{eqnarray}
where
\begin{equation}
  \theta=\arctan\left[2(\tilde{\Omega}_{R}-\tilde{\Delta})/\tilde{A}\right]\label{eq:theta}
\end{equation}
and $|\pm\rangle$ are the bare levels of the TLS: $\sigma_{z}|\pm\rangle=\pm|\pm\rangle$.

By using the above results, we have obtained the time evolution operator as
\begin{equation}
U_{S}(t)=e^{-S(t)}R^{\dagger}(t)e^{-iH_{{\rm CHRW}}^{\prime}t}.
\end{equation}
Provided that the initial states of TLS is $|\psi_{\pm}(0)\rangle=|\widetilde{\pm}\rangle,$
we have the final state at the time $t$ given by
\begin{eqnarray}
|\psi_{\pm}(t)\rangle & = & U_{S}(t)|\psi_{\pm}(0)\rangle\nonumber \\
 & = & e^{\mp i\frac{1}{2}\tilde{\Omega}_{R}t}e^{-S(t)}R^{\dagger}(t)|\widetilde{\pm}\rangle\nonumber \\
 & \equiv & e^{-i\varepsilon_{\pm n}t}|u_{\pm n}(t)\rangle,
\end{eqnarray}
where
\begin{eqnarray}
|u_{\pm n}(t)\rangle & = & e^{i(n+1/2)\omega_{l}t}e^{-S(t)}R^{\dagger}(t)|\widetilde{\pm}\rangle,\label{eq:ut}\\
\varepsilon_{\pm n} & = & (\omega_{l}\pm\tilde{\Omega}_{R})/2+n\omega_{l}.\label{eq:quer}
\end{eqnarray}
It is evident that $|u_{\pm n}(t)\rangle=|u_{\pm n}(t+2\pi/\omega_{l})\rangle$
are periodic in time. According to the Floquet theory, we identify
that $\varepsilon_{\pm n}$ and $|u_{\pm n}(t)\rangle$ are the quasienergies
and Floquet states, respectively. In contrast, it is straightforward
to derive the quasienergies and Floquet states within the rotating-wave approximation (RWA) by replacing
the modified quantities $\tilde{A}$ and $\tilde{\Delta}$ with the
corresponding bare quantities ($A$, $\Delta=\omega_0-\omega_L$), and $e^{-S(t)}$ with $1$ in Eqs.~(\ref{eq:ut})
and (\ref{eq:quer}).

\subsection{Comparison of quasienergy and Floquet state }

\begin{figure}
  \includegraphics[width=12cm]{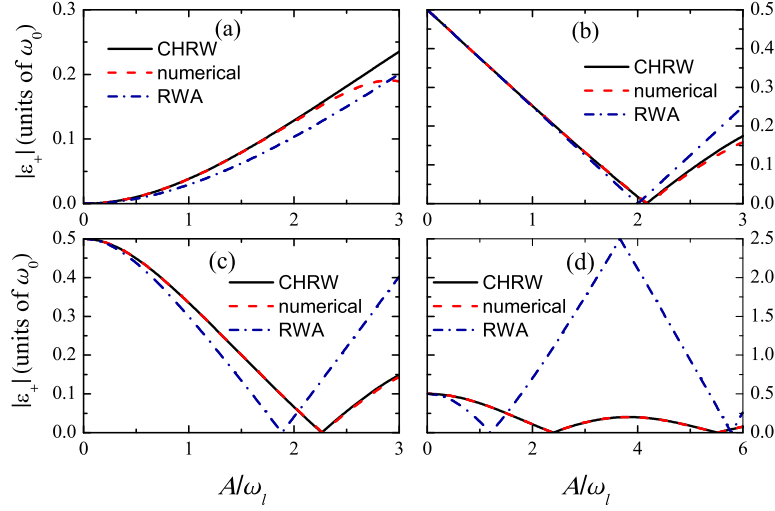}\\
  \caption{(Color online) The absolute value of quasienergy $|\varepsilon_{+}|$ as a function of the ratio $A/\omega_l$ for various driving frequencies: (a) $\omega_l=0.5\omega_0$, (b) $\omega_l=\omega_0$, (c) $\omega_l=1.5\omega_0$, and (d) $\omega_l=5\omega_0$.}\label{fig1}
\end{figure}

We show the validity of our treatment as compared to the numerically exact and the RWA results. First, we
compare the results of quasienergies by our CHRW method, numerically
exact treatment of Floquet Hamiltonian and the RWA. Note that
$(\varepsilon_{+n}+\varepsilon_{-m})\,\mathrm{mod\,}\omega_{l}=0$~\cite{shirley1965solution},
it is sufficient to compare the quasienergy $\varepsilon_{+}$ in the range of $(-\omega_l/2,\omega_l/2]$, whose absolute values are shown in Fig.~\ref{fig1}. Obviously, our results agree well with the numerically exact ones when
$A/\omega_{l}<2$ for both resonance and off-resonance cases. In particular,
for $\omega_{0}/\omega_{l}\ll1$, our method works quite well even
though $A/\omega_{l}\rightarrow6$. However, the RWA is valid only
for the (near-) resonance case and $A/\omega_{l}<1.5$ from the viewpoint of the quasienergies.

Second, we reveal the accuracy of the element between the Floquet states $X_{\alpha\beta,n}^{(j)}$ calculated
by our method. We consider
$\hat{X}^{z}=\frac{\sigma_{z}}{2}$. By using Eqs.~(\ref{eq:xabn})
and~(\ref{eq:ut}), we can obtain the expression for $X_{\alpha\beta,n}^{z}$
with $|u_{\alpha}(t)\rangle\equiv|u_{\alpha0}(t)\rangle$:
\begin{eqnarray}
X_{\alpha\beta,n}^{z} & = & \frac{1}{2}\left\{ c_{\alpha\beta}\left[\delta_{n,0}J_{0}\left(\frac{A\xi}{\omega_{l}}\right)+\sum_{k=1}^{\infty}J_{2k}\left(\frac{A\xi}{\omega_{l}}\right)(\delta_{n,2k}+\delta_{n,-2k})\right]\right.\nonumber \\
 &  & +d_{\alpha\beta}\left[\sum_{k=1}^{\infty}J_{2k-1}\left(\frac{A\xi}{\omega_{l}}\right)(\delta_{n,2k-2}-\delta_{n,-2k})\right]\nonumber \\
 &  & \left.-d_{\beta\alpha}\left[\sum_{k=1}^{\infty}J_{2k-1}\left(\frac{A\xi}{\omega_{l}}\right)(\delta_{n,2k}-\delta_{n,-2k+2})\right]\right\} ,\label{eq:xabne}
\end{eqnarray}
where
\begin{eqnarray}
c_{\alpha\beta} &=&\langle\widetilde{\alpha}|\sigma_z|\widetilde{\beta}\rangle\nonumber\\
& = & \sin(2\theta)(1-\delta_{\alpha,\beta})+(1-2\delta_{\alpha,-})\delta_{\alpha,\beta}\cos(2\theta),\\
d_{\alpha\beta} &=&\langle\widetilde{\alpha}|\sigma_-|\widetilde{\beta}\rangle\nonumber\\
& = & (\sin^{2}\theta-\delta_{\beta,+})(1-\delta_{\alpha,\beta})+\left(\frac{1}{2}-\delta_{\beta,-}\right)\delta_{\alpha,\beta}\sin(2\theta).
\end{eqnarray}
The derivation of the coefficient $X^{(j)}_{\alpha\beta,n}$ is given in the Appendix.
In Figs.~\ref{fig2} and \ref{fig3}, we show the behaviors of $|X_{++,0}^{z}|$
and $|X_{++,2}^{z}|$ as a function of ratio $A/\omega_{l}$,
respectively. We find that results of the CHRW method are in good agreement with numerical
results when $A/\omega_{l}<2$ for both off- and on-resonance cases.
Figure.~\ref{fig3} shows that $X_{\alpha\beta,n}^{z}$
for $n\neq0$ is significantly enhanced as $A$ increases. In contrast, the RWA results is $X_{\alpha\beta,n}^{z} \equiv 0$ which is totally different from $X_{\alpha\beta,n}^{z}\neq0$ for $n\neq0$ obtained by the exact and CHRW methods. It indicates that the incorrect RWA coefficient $X^{(j)}_{\alpha\beta,n}$ for $n \neq 0$ can not predict the accurate dynamics and analysis the features of the spectrum in the strong driving case.

From the above comparison, we find that our approach can give the correct result with high accuracy in comparison with the numerically exact results when $ A/\omega_{l}<2$ (or $A/\omega_{l}<6$ and $\omega_{0}/\omega_{l}\ll1$). On the contrary, the RWA is valid
only when $A/\omega_{l}\ll1$. It is therefore reasonable to apply Floquet states (\ref{eq:ut}) and associated quasienergies (\ref{eq:quer}) obtained by the CHEW method
to analytically evaluate the fluorescence spectrum.

\subsection{Standard and modified Mollow triplets}

In order to give the fluorescence spectrum of the model (\ref{eq:model}),
it requires us to calculate $X_{\alpha\beta,n}^{x}$, $X_{\alpha\beta,n}^{+}$,
and $\gamma_{\alpha\beta,n}^{(j)\pm}$ with $j=x,z$, which reads
\begin{eqnarray}
X_{\alpha\beta,n}^{x} & = & \frac{1}{2}(d_{\alpha\beta}\delta_{n,-1}+d_{\beta\alpha}\delta_{n,1}),\\
X_{\alpha\beta,n}^{+} & = & \frac{1}{2}\left\{ d_{\beta\alpha}\left[1+J_{0}\left(\frac{A\xi}{\omega_{l}}\right)\right]\delta_{n,1}+d_{\beta\alpha}\sum_{k=1}^{\infty}J_{2k}\left(\frac{A\xi}{\omega_{l}}\right)(\delta_{n,2k+1}+\delta_{n,-2k+1})\right.\nonumber \\
 &  & +d_{\alpha\beta}\left[1-J_{0}\left(\frac{A\xi}{\omega_{l}}\right)\right]\delta_{n,-1}-d_{\alpha\beta}\sum_{k=1}^{\infty}J_{2k}\left(\frac{A\xi}{\omega_{l}}\right)(\delta_{n,-2k+1}+\delta_{n,-2k-1})\nonumber \\
 &  & \left.-c_{\alpha\beta}\sum_{k=1}^{\infty}J_{2k-1}\left(\frac{A\xi}{\omega_{l}}\right)(\delta_{n,2k-1}-\delta_{n,-2k+1})\right\} ,\label{eq:xabp}\\
\gamma_{\alpha\beta,n}^{(j)+} & = & \pi G^{(j)}(-\Delta_{\alpha\beta,n})-iR^{(j)}(-\Delta_{\alpha\beta,n})=(\gamma_{\beta\alpha,-n}^{(j)-})^{\ast},
\end{eqnarray}
where $G^{x}(\omega)=\sum_{{\bf k}}g_{{\bf k}}^{2}\delta(\omega_{{\bf k}}-\omega)$
and $G^{z}(\omega)=\sum_{{\bf q}}f_{{\bf q}}^{2}\delta(\nu_{{\bf q}}-\omega)$
are the spectral functions of the electromagnetic reservoir and
dephasing reservoir, respectively. In this work, we consider that
the electromagnetic reservoir is of broadband type with $G^{x}(\omega)=\frac{2}{\pi}\kappa$ where $\kappa$ is the radiative decay rate,
and the dephasing bath is of Ohmic type with $G^{z}(\omega)=\alpha\omega e^{-\omega/\omega_{c}}$ where $\alpha$ is the dimensionless
coupling strength and $\omega_{c}$ is the cut-off frequency. Provided that the reservoirs are at zero temperature and in vacuum states, the
basic ingredients in the FBM master equation can be solved
\begin{eqnarray}
W_{-+} & = & 2\pi\sum_{n}\sum_{j=x,z}|X_{+-,n}^{(j)}|^{2}G^{(j)}(\Delta_{-+,n}),\label{eq:newW}\\
\gamma_{{\rm rel}} & = & 2\pi\sum_{n}\sum_{j=x,z}|X_{+-,n}^{(j)}|^{2}[G^{(j)}(\Delta_{+-,n})+G^{(j)}(-\Delta_{+-,n})],\\
\gamma_{{\rm deph}} & = & \pi\sum_{n}\sum_{j=x,z}\big\{|X_{-+,n}^{(j)}|^{2}[G^{(j)}(\Delta_{-+,n})+G^{(j)}(-\Delta_{-+,n})]\nonumber \\
 &  & +2|X_{++,n}^{(j)}|^{2}[G^{(j)}(n\omega_{l})+G^{(j)}(-n\omega_{l})]\big\}.\label{eq:rdephn}
\end{eqnarray}
Therefore, we determine the fluorescence
spectrum of the driven TLS interacting with two reservoirs in the following.

\subsubsection{The driving-induced asymmetry}

First, we reveal that how the standard Mollow triplet recovers from
Eq.~(\ref{eq:sfun}) in the absence of the dephasing reservoir ($f_{\bf q} = 0$). We
consider the coefficient $X_{\alpha\beta,n}^{+}$ within the RWA,
\begin{eqnarray}
X_{\alpha\beta,n}^{+} & = & d_{\beta\alpha}^{(\mathrm{RWA})}\delta_{n,1},
\end{eqnarray}
where $d_{\beta\alpha}^{(\mathrm{RWA})}$ is given by $d_{\beta\alpha}$
with $\theta_{\rm{A}}=\arctan[2(\Omega_{R}-\Delta)/A]$. Here $\Delta=\omega_{0}-\omega_{l}$
is the bare detuning and $\Omega_{R}=\sqrt{\Delta^{2}+A^{2}/4}$ is
the Rabi frequency. For $X_{\alpha\beta,n\neq1}^{+}=0$,  the four components of the fluorescence spectra (the delta-function and three Lorentzians) associated with $n=1$ survive in Eq.~(\ref{eq:sfun}). To examine the spectral features, we need to calculate the Floquet-state population $\rho^{\rm{ss}}_{\alpha\alpha}$, relaxation rate $\Gamma_{\rm{rel}}$, and dephasing rate $\Gamma_{\rm{deph}}$, which is easily evaluated from Eqs.~(\ref{eq:newW})-(\ref{eq:rdephn}) and given by
\begin{eqnarray}
\rho_{++}^{{\rm ss}} & = & \frac{\sin^{4}\theta_{\mathrm{A}}}{\cos^{4}\theta_{\mathrm{A}}+\sin^{4}\theta_{\mathrm{A}}}=1-\rho_{--}^{{\rm ss}},\\
\gamma_{{\rm rel}} & = & \kappa(\sin^{4}\theta_{\mathrm{A}}+\cos^{4}\theta_{\mathrm{A}}),\\
\gamma_{{\rm deph}} & = & \frac{\kappa}{2}[\sin^{4}\theta_{\mathrm{A}}+\cos^{4}\theta_{\mathrm{A}}+\sin^{2}(2\theta_{{\rm A}})].
\end{eqnarray}
It is straightforward to verify that the equality $|X^+_{+-,1}|^2\rho_{++}^{\rm{ss}}=|X^+_{-+,1}|^2\rho_{--}^{\rm{ss}}$ exactly holds within the RWA, which is known as the detailed balance condition. It guarantees that the spectrum is always symmetrical with respect to the center~\cite{cohen1992atom}.

In particular, when $\Delta=0$,
we have $\theta_{\rm{A}}=\pi/4$, which leads to $|X_{\alpha\beta,1}^{+}|=\frac{1}{2}$, $\gamma_{{\rm rel}}=\kappa/2$,
$\gamma_{{\rm deph}}=3\kappa/4$, and $\rho_{++}^{{\rm ss}}=\rho_{--}^{{\rm ss}}=1/2$.
Thus, our spectrum Eq.~(\ref{eq:sfun}) recovers the standard incoherent Mollow triplet:
\begin{equation}
I(\omega)\propto\frac{\kappa/8}{(\omega-\omega_{l})^{2}+\kappa^{2}/4}+\frac{3\kappa/32}{(\omega-\omega_{l}-\Omega_{R})^{2}+9\kappa^{2}/16}+\frac{3\kappa/32}{(\omega-\omega_{l}+\Omega_{R})^{2}+9\kappa^{2}/16}.
\end{equation}
Here we find that the coherent component (the delta function) vanishes. This arises because
of the secular approximation which omits the terms of order $\kappa/\Omega_{R}$
and higher.

Second, we demonstrate the spectral properties beyond the RWA and
in the absence of the dephasing reservoir. The results are as follows: (i) the positions of center components in the Mollow triplets appear at odd multiple of the fundamental driving frequency. By Eq.~(\ref{eq:xabp}), we get $|X_{\alpha\beta,n}^{+}|=0$ for $|n|=0,2,4,\cdots$,
but $|X_{\alpha\beta,n}^{+}|\neq0$ for $|n|=1,3,5,\cdots$. A nonzero $|X_{\alpha\beta,n}^{+}|$ means that the transition $|u_{\alpha n}(t)\rangle\rightarrow|u_\beta(t)\rangle$ is allowed when $n$ is odd, leading to the generation of Mollow
triplets centered at odd multiples of driving frequency $n\omega_{l}$ $(|n|=1,3,5,\cdots)$. Actually,
the components centered at negative frequencies give negligible contributions
to $I(\omega>0)$. The main ingredients of fluorescence spectrum are
the emission lines at positive frequencies; (ii) the two sidebands around $\omega_{l}$ possess
unequal intensities in the case of the strong harmonic driving. This type of asymmetry has been proved in previous
works~\cite{yan2013effects,browne2000resonance}. It turns out here the asymmetry of the first-order triplet can
be understood by the inequality $|X_{-+,1}^{+}|^{2}\rho_{--}^{{\rm ss}} < |X_{+-,1}^{+}|^{2}\rho_{++}^{{\rm ss}}$,
which means the violation of the detailed balance condition.

We examine analytically the violation of the detailed balance condition. Similar to the RWA case, we have the steady Floquet-state population $\rho_{++}^{{\rm ss}}=\frac{\sin^{4}\theta}{\cos^{4}\theta+\sin^{4}\theta}=1-\rho_{--}^{{\rm ss}}$ while $\theta$ is given by Eq.~(\ref{eq:theta}). On the other hand, we find  $|X^+_{-+,1}|=\frac{1}{2}\left|\sin^{2}\theta\left[1+J_{0}\left(\frac{A\xi}{\omega_{l}}\right)\right]+J_{2}\left(\frac{A\xi}{\omega_{l}}\right)\cos^{2}\theta-J_{1}\left(\frac{A\xi}{\omega_{l}}\right)\sin(2\theta)\right|$ and $|X_{+-,1}^{+}|=\frac{1}{2}\left|\cos^{2}\theta\left[1+J_{0}\left(\frac{A\xi}{\omega_{l}}\right)\right]+J_{2}\left(\frac{A\xi}{\omega_{l}}\right)\sin^{2}\theta+J_{1}\left(\frac{A\xi}{\omega_{l}}\right)\sin(2\theta)\right|$.
Provided that $A/\omega_l\ll1$ and $J_{n}\left(\frac{A\xi}{\omega_{l}}\right)$ with $n\geq1$ is omitted, we recover the detailed balance condition. However, when $A/\omega_l$ are large enough, the contributions from higher-order Bessel function  $J_n\left(\frac{A\xi}{\omega_{l}}\right)$ become important. In Fig.~\ref{fig4}(a), we show the process of the violation of the detailed balance condition with the increase
of $A$. We find that the condition is apparently violated when $A>0.2\omega_{0}$ for $\omega_{l}=\omega_{0}$, which differs from the prediction
of $A>0.8\omega_{0}$ given in Ref.~\cite{browne2000resonance} (their notation $\Omega$ is equal to $2A$). In addition, Fig.~\ref{fig4}(b) shows that $|X_{-+,1}^{+}|^{2}\rho_{--}^{{\rm ss}}<|X_{+-,1}^{+}|^{2}\rho_{++}^{{\rm ss}}$ holds for both the on- and off-resonance cases in the strong-driving regime. As a result, we find that the red sideband is suppressed while the blue one is enhanced with the increase of $A$. In contrast, we notice that this property can not be given from the traditional quantum optical master equation~\cite{ho1986floquet} (see Figs.~8 and~12 of the reference).

From the above discussion, we notice that we can gain an insight into the spectral properties of the fluorescence by analyzing
the properties of $X_{\alpha\beta,n}^{+}$ as well as the steady Floquet-state populations. In particular, $X_{\alpha\beta,n}^{+}$ is uniquely determined by the driving. However, the populations are generally influenced by both the driving and the reservoirs. In the following, we
demonstrate the effect of the pure dephasing coupling on the spectrum.

\subsubsection{The depahsing-induced asymmetry}
We explore the effect of the pure dephasing reservoir ($f_q \neq 0$) on the spectrum.
In Fig.~\ref{fig5}, we show the fluorescence spectra for the on- and off-resonance cases.
It is found that the spectra exhibit asymmetry with the increase of $\alpha$ which characterizes the strength of the dephasing coupling.
It is the dephasing coupling that results in the enhancement of the red sideband and the suppression of the blue one,
which is opposed to the feature of the harmonic-driving-induced asymmetry (the suppression of the red sideband and enhancement of the blue sideband)~\cite{yan2013effects}.
Moreover, the dephasing-induced asymmetry becomes more apparent for the blue detuned driving [see Fig.~\ref{fig5}(c)] than the resonant driving [see Fig.~\ref{fig5}(a)]. This property is further illustrated in Fig.~\ref{fig6}. We find that the difference in the weights of the two sidebands becomes evident for both RWA and non-RWA driving with the increase of $\omega_l$ for fixed $A$.

Figure~\ref{fig5}(b) shows a phenomenon of the interplay between harmonic-driving and dephasing induced asymmetry.
For a certain harmonic driving strength, it is possible to realize the inequality $|X_{-+,1}^{+}|^{2}<|X_{+-,1}^{+}|^{2}$ resulting in the spectral feature with a suppressed red sideband and an enhanced blue sideband. For $\alpha=0$, the spectral asymmetry is only determined by the harmonic driving, i.e. the blue sideband is higher than the red one. When $\alpha$ increases, the blue sideband comes to be suppressed while the red one becomes enhanced. This means that the spectral asymmetry is dominated by the dephasing coupling. In Fig.~\ref{fig6}(a), it is clear to see the CHRW method gives the crossover of the two types of asymmetry induced by harmonic driving and dephasing coupling. When $\omega_l<0.97\omega_0$, the red sideband is lower than the blue one, which is the feature of the driving-induced asymmetry. While $\omega_l>0.97\omega_0$, the red sideband becomes higher than the blue one, which is the feature of the dephasing-induced asymmetry. In contrast, the red sideband of the RWA is always higher than the blue one because the RWA does not take into account the effects of the counter-rotating driving term, which means that the RWA spectrum are asymmetric. In Fig.~\ref{fig6}(b), we show the spectrum for the non-RWA and RWA cases for $\omega_l=0.97\omega_0$. Interestingly, the non-RWA spectrum is symmetric while the RWA spectrum is asymmetric.

We explore the effects of the dephasing reservoir by analyzing the population $\rho^{\rm{ss}}_{\alpha\alpha}$ which influences the weights of each components of the fluorescence and depends on the properties of the reservoir.
In Figs.~\ref{fig7}(a) and~\ref{fig7}(b), we show the population $\rho^{\rm{ss}}_{++}$ as functions of $\omega_l$ and $A$ for both RWA and harmonic driving,
respectively. It is evident that $\rho_{++}^{\rm{ss}}$ becomes smaller for the dephasing coupling $\alpha>0$ than for $\alpha=0$ [the solid (dot-dashed) line with $\alpha=0.01$ is always below the dashed (dotted) line with $\alpha=0$]. Moreover, as $\omega_l$ increases, $\rho_{++}^{\rm{ss}}$ becomes smaller for $\alpha>0$ than for $\alpha=0$. It is feasible to understand the role of the dephasing coupling on the spectra by examining $\rho^{\rm{ss}}_{++}$ of the RWA driving case. We obtain the expression for $\rho^{\rm{ss}}_{++}$ in the RWA:
\begin{equation}
\rho_{++}^{{\rm ss}}(\alpha)=\frac{\kappa\sin^{4}\theta_{{\rm A}}}{\kappa(\cos^{4}\theta_{{\rm A}}+\sin^{4}\theta_{{\rm A}})+\frac{\pi}{2}\alpha G^{z}(\Omega_{R})\sin^{2}(2\theta_{{\rm A}})}.\label{eq:rho11srwa}
\end{equation}
It follows from Eq.~(\ref{eq:rho11srwa}) that $\rho_{++}^{{\rm ss}}(\alpha)\leq\rho_{++}^{{\rm ss}}(0)$, i.e., the steady population $\rho_{++}^{{\rm ss}}$ decreases as $\alpha$ increases. In comparison with $|X^+_{+-,1}|^2\rho_{++}^{\rm{ss}}(0)=|X^+_{-+,1}|^2\rho_{--}^{\rm{ss}}(0)$ for $\alpha=0$, we conclude that $|X^+_{+-,1}|^2\rho_{++}^{\rm{ss}}(\alpha)<|X^+_{-+,1}|^2\rho_{--}^{\rm{ss}}(\alpha)$ for $\alpha>0$. This inequality leads to the generation of the enhanced red sideband and suppressed blue sideband, which is qualitatively consistent with the experiment observation~\cite{ulhaq2013detuning}. The similar discussion can be explored in the harmonic driving case. Therefore, the dephasing induced asymmetry can be attributed to the modifications to the steady Floquet-state populations caused by the dephasing reservoir.

\section{Application to the biharmonic driving}\label{appli2}
In order to show the high efficiency and the advantage of our formalism Eqs.~(\ref{eq:H0}) to (\ref{eq:sfun}) to any periodic driving, we study the biharmonic driving case with the aid of numerical method. The biharmonically driven TLS is described by the Hamiltonian:
\begin{equation}
  H_S(t)=\frac{1}{2}\omega_0\sigma_z+\frac{A}{2}[\cos(\omega_l t)+r\cos(2\omega_l t+\phi)]\sigma_x,
\end{equation}
where $\phi$ is the relative phase of the signals and $r$ is the
relative amplitude. Similar to the harmonic driving case, the key task is to evaluate the Floquet states and quasienergies for the Floquet Hamiltonian, which can be done by the numerical treatment.

Since $|u_\alpha(t)\rangle$ is periodic in time, we can formally expand the Floquet state as $|u_\alpha(t)\rangle=\sum_n u^{(\alpha)}_{\gamma n}e^{in\omega_l t}|\gamma\rangle$ where $\gamma$ is the index of the two levels and $n$ is an integer. Substituting this expansion to Eq.~(\ref{eq:eigeneq}), one finds that
the Fourier coefficients satisfy the following equation
\begin{equation}
\sum_{n,\gamma}\mathcal{H}_{\delta m,\gamma n}u_{\gamma n}^{(\alpha)}=\varepsilon_{\alpha}u_{\delta m}^{(\alpha)},\label{eq:eigensol}
\end{equation}
where $\mathcal{H}_{\delta m,\gamma n}=\frac{\omega_{l}}{2\pi}\int_{0}^{2\pi/\omega_{l}}\langle\delta|e^{-im\omega_{l}t}[H_{S}(t)-i\partial_{t}]e^{in\omega_{l}t}|\gamma\rangle dt\equiv\left\langle \delta,m|\mathcal{H}_{F}|\gamma,n\right\rangle$ is the element of Floquet Hamiltonian in the Sambe space~\cite{sambe1973steady}. The Sambe space is a composite Hilbert space spanned by the basis $\{|\gamma,n\rangle,n\in\mathbb{Z};\langle t|n\rangle=\exp(in\omega_l t)\}$. Thus, the time-dependent Eq.~(\ref{eq:eigeneq}) is now converted into a time-independent equation~(\ref{eq:eigensol}), which is an eigenvalue problem: $\mathcal{H}_F|u^{(\alpha)}\rangle=\varepsilon_\alpha|u^{(\alpha)}\rangle$ where $|u^{(\alpha)}\rangle=\sum_{n,\gamma}u^{(\alpha)}_{\gamma n}|\gamma,n\rangle$. For the biharmonically driven TLS, we readily obtain the matrix form of Floquet Hamiltonian as follows:
\begin{equation}
\mathcal{H}_{F}=\left(\begin{array}{c|cccccc|c}
\ddots& |-,-1\rangle & |+,-1\rangle & |-,0\rangle & |+,0\rangle & |-,1\rangle & |+,1\rangle & \\ \hline
|-,-1\rangle & -\frac{\omega_{0}}{2}-\omega_{l} & 0 & 0 & \frac{A}{4} & 0 & \frac{rA}{4}e^{i\phi}\\
|+,-1\rangle & 0 & \frac{\omega_{0}}{2}-\omega_{l} & \frac{A}{4} & 0 & \frac{rA}{4}e^{i\phi} & 0\\
|-,0\rangle & 0 & \frac{A}{4} & -\frac{\omega_{0}}{2} & 0 & 0 & \frac{A}{4}\\
|+,0\rangle & \frac{A}{4} & 0 & 0 & \frac{\omega_{0}}{2} & \frac{A}{4} & 0\\
|-,1\rangle & 0 & \frac{rA}{4}e^{-i\phi} & 0 & \frac{A}{4} & -\frac{\omega_{0}}{2}+\omega_{l} & 0\\
|+,1\rangle & \frac{rA}{4}e^{-i\phi} & 0 & \frac{A}{4} & 0 & 0 & \frac{\omega_{0}}{2}+\omega_{l}\\ \hline
 &  &  &  &  &  &  & \ddots
\end{array}\right).
\end{equation}
By introducing an appropriate truncation to the matrix, we can numerically diagonalize $\mathcal{H}_F$ and simultaneously obtain its eigenvalues and eigenstates $|u^{(\alpha)}\rangle$. The eigenvalue is actually the quasienergy. The eigenstate leads to the Floquet state by the relations $|u_\alpha(t)\rangle=\langle t|u^{(\alpha)}\rangle$ and $\langle t|n\rangle=\exp(in\omega_l t)$.
After obtaining the quasienergies and Floquet states, we can determine the required quantity $X^{(j)}_{\alpha\beta,n}$ of fluorescence spectrum~(\ref{eq:sfun}). In principle, the quantity can be formally evaluated as follows:
\begin{eqnarray}
X_{\alpha\beta,n}^{(j)} & = & \frac{\omega_{l}}{2\pi}\int_{0}^{2\pi/\omega_{l}}\langle u_{\alpha}(t)|\hat{X}^{(j)}|u_{\beta}(t)\rangle e^{-in\omega_{l}t}dt\nonumber \\
 & = & \frac{\omega_{l}}{2\pi}\int_{0}^{2\pi/\omega_{l}}\langle u_{\alpha n}(t)|\hat{X}^{(j)}|u_{\beta}(t)\rangle dt\nonumber \\
 & = & \sum_{l,\gamma,\delta}\left[u_{\gamma l-n}^{(\alpha)}\right]^{\ast}u_{\delta l}^{(\beta)} \langle \gamma|\hat{X}^{(j)}|\delta \rangle.
\end{eqnarray}
Then, we can calculate the steady Floquet-states populations, relaxation and dephasing rates by using the same procedure as the former section, which completely determines the fluorescence spectrum. It is evident that the key task is to diagonalize the Floquet Hamiltonian numerically and obtain the Floquet states and quasienergies. In contrast, the Floquet states and quasienergies of harmonic driving case in the former section are analytically derived based on the unitary transformation.

We now illustrate how the second component of the biharmonic field modify the spectrum. In Fig.~\ref{fig8}(a), we show the fluorescence spectra for the different relative amplitudes in the case $\omega_l=\omega_0$. The spectra with $r\neq 0$ show more intensities of higher-order Mollow triplets centered at both even and odd multiples of the driving frequency, which qualitatively differs from the the spectrum with $r=0$. Note that the harmonic driving of frequency $2\omega_0$ induces the triplets centered at $2n\omega_0$ with $n=1,3,5,\cdots$, which means that the second component ($2\omega_l$) of the biharmonic field is responsible for the generation of the triplets centered at $2\omega_0$, $6\omega_0$, $10\omega_0$,$\cdots$. However, we find that in the plot there is an additional triplet centered at $4\omega_0$. It indicates that the higher-order triplets are not a simple superposition of the independent spectra of each harmonic driving mode $\omega_l$ and $2\omega_l$. It is because that some forbidden transition channels of Floquet states for single harmonic driving case are permitted in the presence of biharmonic field. In Fig.~\ref{fig8}(b), we show the behaviors of $X_{+-,n}$ for $n=1,2,3,4$ as a function of $r$, shown. It is clear that $|X_{+-,n}|$ for $n=2,4$ increases from zero with the increase of $r$. It means that the transition channels forbidden in the harmonic case are turned on in the biharmonic case.
Moreover, the coefficients $X_{+-,n}$ with $n=2,3,4$ are enhanced with the increase of $r$. It leads to the enhanced transition-related factor and hence we see more intense higher-order triplets centered at odd and even multiples of driving frequency for the biharmonic driving.

We demonstrate the effect of relative phase $\phi$ on the spectra. In Fig.~\ref{fig9}(a), we show the spectra for various relative phases. The results indicate that the relative phase just influences the line shapes of certain higher-order triplets [see the inset of Fig.~\ref{fig9}(a)]. To examine the effect of the phase,
in reality, we could numerically calculate the population $\rho^{\rm{ss}}_{\alpha\alpha}$ and the coefficient $X^+_{\alpha\beta,n}$ as a function of the relative phase $\phi$. One can verify that the steady Floquet-state populations are almost independent of the phase but some $X^+_{\alpha\beta,n}$ are sensitively dependent on the phase.  In Fig.~\ref{fig9}(b), we show the behavior of $|X^+_{+-,n}|$ for $n=1,2,3,4$ with the variation of $\phi$ when $\omega_l=\omega_0$. The coefficients $|X^+_{+-,n}|$ for $n=1,2,$ and $4$ do not change with the increase of phase. However, the coefficients $|X_{+-,n}|$ for $n=3$ does vary with the phase. In particular, it is the phase dependence of $|X^+_{\alpha\beta,3}|$ that leads to the phase-dependent triplet centered at $3\omega_0$ in contrast to the phase-independent Mollow triplets centered at $\omega_0$ and $2\omega_0$.

The present formalism can also be applied to explore the multiple resonance induced fluorescence of the biharmonically driven TLS, which also shows the advantage of the present formalism. First, we need to evaluate the positions where multiple resonance occurs. The evaluation can be carried out easily with Floquet theory by calculating the time-averaged transition probability similar to Shirley's origin work~\cite{shirley1965solution}. The mean transition probability from $|-\rangle$ to $|+\rangle$ is given by~\cite{shirley1965solution}
\begin{equation}
  \bar{P}=\frac{1}{2}(1-4|X^z_{++,0}|^2).
\end{equation}
The maximum of $\bar{P}$ indicates the emergence of the resonance. In Fig.~\ref{fig10}, we show the $\bar{P}$ as a function of $\omega_l$ for the fixed driving strength $A=0.5\omega_0$ and $\phi=0$. When $r=1$, it is evident that a series of peaks emerge, indicating a series of resonance, which correspond to $\omega_l/\omega_0=0.9933$, $0.5572$, $0.3844$, $0.2834$, $\cdots$. The first two frequencies  $\omega_l/\omega_0=0.9933$, and $0.5572$ corresponds to the main resonance frequencies of the two modes of the biharmonic field. The other frequencies are the multiple resonance frequencies. When $r=0$, we find two resonance peaks in the considered frequency range. In particular, the width of multiple resonance for harmonic driving is much narrower than those of the biharmonic driving.

We discuss the difference between non-RWA and RWA theory in the biharmonic driving case. In Fig.~\ref{fig10}, we also provide the results of $\bar{P}$ calculated from the RWA Hamiltonian
\begin{equation}
  H_{\mathrm{RWA}}(t)=\frac{1}{2}\omega_0\sigma_z+\frac{A}{4}\left[(e^{i\omega_l t}+r e^{i2\omega_l t})\sigma_-+\mathrm{h.c.}\right].
\end{equation}
The RWA has been used by Ficek and Freedhoff to study the biharmonic driving with two incommensurate frequencies~\cite{ficek1993resonance}. By comparison of non-RWA and RWA results, we find the Bloch-Siegert shift which is the non-RWA resonance peaks shift from the RWA ones. Moreover, the RWA generally leads to the unfaithful width of resonance. This is apparent in the multiple resonance.

Figure~\ref{fig11} shows the resonance fluorescence of the biharmonically driven TLS under the resonance conditions. It is evident that the most intense fluorescence occurs around $n\omega_l$ $(n=1,2,3,4,\cdots)$ for each resonance. We first concentrate on the features of non-RWA spectra. For the first main resonance ($\omega_l=0.9933\omega_0$), we find that the spectrum is similar to that of the harmonic driving case. For the second main resonance ($\omega_l=0.5572\omega_0$), the spectrum has two observable triplets induced by the two modes of the biharmonic field. For multiple resonance [Figs.~\ref{fig11}(c) and~\ref{fig11}(d)], the spectra generally has multi-peak structure. Besides, the splitting of the triplet decreases because the resonance width decreases rapidly for multiple resonance. By comparison, it is obvious that the structure of the non-RWA spectra can be significantly different from the RWA one, in particular, for the multiple resonance. Therefore, it turns out that the RWA is invalid for exploring the multiple resonance under the strong biharmonic driving.

We further consider the effect of the phase under the multiple resonance condition. Figure~\ref{fig12}(a) shows the influence of the phase on the multiple resonance induced fluorescence. In contrast to Fig.~\ref{fig9}(a) with $\omega_l=\omega_0$, it is clear to see that all the three triplets are varied as the change of phase. In Fig.~\ref{fig12}(b), we show the behavior of $|X^+_{+-,n}|$ with $n=1,2,3,4$ when $\omega_l=0.3844\omega_0$ and $A=0.5\omega_0$. It is obvious to see that $X^+_{+-,n}$ with $n=1,2,3,4$ is modulated by the phase. This indicates that the line shape of fluorescence could be modulated by the phase under multiple resonance condition.

From the analysis above, we find that even though the analytical expressions for the Floquet states and quasienergies are not available in the complicated driving case, one can still easily calculate the fluorescence spectrum and analyze the spectral features according to Eq.~(\ref{eq:sfun}). The difference of the spectral features between the biharmonic and harmonic driving cases can be mainly attributed to the quantity $X^+_{\alpha\beta,n}$, which is corresponding to the transition of the Floquet states and is uniquely determined by the properties of the driving field. In addition, the present formalism also allows us to study the multiple resonance induced spectrum.

\section{Application to the multiharmonic driving}\label{appli3}

After applying the present formalism to the harmonic and biharmonic driving signals, we will show the advantage of the formalism in dealing with complicated driving signal, such as multiharmonically periodic driving, with the aid of the numerical method.
We consider that the TLS is driven by the square-wave (SW) like driving:
\begin{equation}
  H_S(t)=\frac{1}{2}\omega_0\sigma_z+\frac{A}{2}\sigma_x\sum_{l=1}^{N}\frac{\sin[(2l-1)\omega_l t]}{2l-1}.
\end{equation}
Here $N$ is set for $100$. Using numerical treatment of the Floquet Hamiltonian, we obtain the numerically exact Floquet states and quasienergies for the multiharmonically driven TLS. Once again it is straightforward to calculate the spectrum according to Eq.~(\ref{eq:sfun}). We state that the procedure of calculating the spectrum is the same as the former case. The spectra for the SW-like driving are shown in Fig.~\ref{fig13}. Surprisingly, we can hardly distinguish the first-order triplet centered at $\omega_l$ of harmonic driving from that of SW-like driving. In the comparison with the spectra of the harmonic driving, the higher-order triplets under the SW-like driving have stronger intensity. It turns out that a driving signal consisting of odd multiple of frequency $\omega_l$ can just induce the triplets centered at $n\omega_l$ with $n$ being odd integer. In principle, with Eq.~(\ref{eq:sfun}) at hand, we can understand the spectral features by the similar analysis as we have done in the former sections.

\section{conclusion}
In summary, we have presented the formalism for fluorescence spectrum of a periodically driven TLS based on Floquet-Born-Markov master equation which can be generally extended to treat the case with arbitrarily periodic driving and weak-coupling multiple reservoirs. We show that in the our formalism the driving can be treated exactly according to the Floquet theory while the versatile TLS-reservoir couplings are taken into account up to the second order by the Born-Markov approximation. In the secular limit, we have given an analytical result of fluorescence spectrum with clear physical significance. When the combined Hamiltonian of the TLS and periodic driving is analytically solved, the spectrum is calculated directly. While the Hamiltonian is complex and difficult to be solved analytically, the numerical method are employed to calculate the spectrum.

We have applied the formalism to calculate the fluorescence spectrum of many significant cases and comprehensively explore the spectral features.

(i) We study the fluorescence emitting from the harmonically driven TLS weakly coupled to a radiative reservoir and a dephasing reservoir. By the unitary transformation, we derived analytically the Floquet states and quasienergies for the harmonically driven TLS, which is nearly the same as the numerically exact results over a wide range of the driving parameter space and provide the basis for understanding the spectral features. First, without the dephasing coupling, the general formalism recovers the symmetrical Mollow triplet provided that the RWA of the harmonic driving is introduced. The symmetry of the spectrum is explained by the detailed balance condition. Without the RWA of the driving, the prominent asymmetry of the spectrum can be clarified by the violation of the condition.
Second, we studied the effect of the dephasing coupling on the spectrum.
The dephasing coupling results in the asymmetric Mollow triplet with the enhancement of the red sideband and the suppression of the blue sideband,
which is qualitatively consistent with the experiment observations.
This dephasing-induced asymmetry can be attributed to the change of the steady Floquet-state population resulting from the dephasing coupling.
Moreover, We demonstrate that the interplay between the harmonic driving and dephasing coupling results in different spectral line shapes.

(ii) Apart from the harmonic driving, we applied the formalism to the biharmonic driving case. The biharmonic driving leads to the higher-order Mollow triplets centered at both even and odd multiples of the driving frequency,
which is qualitatively different from the spectra of harmonic driving. Besides, it was found that the relative phase of the driving signals can change the certain higher-order triplet depending on the driving frequency. We also apply the formalism to study the multiple resonance induced resonance fluorescence. We find that the non-RWA and RWA spectra significantly differ from each other under the multiple resonance conditions.

(iii) We applied the formalism to the multiharmonic driving similar to the square-wave signal based on the numerical method. For the moderately intense driving strength, the first-order Mollow triplet under the SW-like driving is almost the same as that of the harmonic driving but the higher-order triplets are more intense than those of the harmonic driving.

All in all, the present formalism provides a unified description for the fluorescence spectrum and is applicable to the situation with complicated periodic driving signals and weak-coupling multiple reservoirs.

\begin{acknowledgments}
The authors thank Prof.~Z. Ficek for valuable discussion. This work was supported by the National Natural Science Foundation of China (Grants No.~11174198, No.~11374208, No.~11581240311, and No.~11474200) and the National Basic Research Program of China (Grant No.~2011CB922202). The work was partially supported by the Shanghai Jiao Tong University SMC-Youth Foundation.
\end{acknowledgments}

\appendix
\section{The derivation of coefficient $X^{(j)}_{\alpha\beta,n}$ for the harmonically driven TLS}
By making use of Eqs.~(\ref{eq:xabn}) and~(\ref{eq:ut}), we can rewrite the expression for $X^{(j)}_{\alpha\beta,n}$ as follows:
\begin{eqnarray}
X_{\alpha\beta,n}^{(j)} & = & \frac{\omega_{l}}{2\pi}\int_{0}^{2\pi/\omega_{l}}\langle u_{\alpha}(t)|\hat{X}^{(j)}|u_{\beta}(t)\rangle e^{-in\omega_{l}t}dt\nonumber \\
 & = & \frac{\omega_{l}}{2\pi}\int_{0}^{2\pi/\omega_{l}}\langle\widetilde{\alpha}|R(t)e^{S(t)}\hat{X}^{(j)}e^{-S(t)}R^{\dagger}(t)|\widetilde{\beta}\rangle e^{-in\omega_{l}t}dt.\label{eq:A1}
\end{eqnarray}
Since $\hat{X}^{(j)}$ may be Pauli matrices and their combination, we have the formal decomposition
\begin{equation}
  R(t)e^{S(t)}\hat{X}^{(j)}e^{-S(t)}R^{\dagger}(t)=\frac{1}{2}C_{j}^{z}(t)\sigma_{z}+C_{j}^{+}(t)\sigma_{-}+C_{j}^{-}(t)\sigma_{+},\label{eq:A2}
\end{equation}
where the time-dependent coefficient $C_j^{\lambda}(t)$ $(\lambda=z,\pm)$ is simply given by
\begin{equation}
  C_j^\lambda(t)=\mathrm{Tr}[\sigma_\lambda R(t)e^{S(t)}\hat{X}^{(j)}e^{-S(t)}R^{\dagger}(t)].
\end{equation}
By substituting Eq.~(\ref{eq:A2}) into Eq.~(\ref{eq:A1}), we arrive at
\begin{eqnarray}
X_{\alpha\beta,n}^{(j)} & = & \langle\widetilde{\alpha}|\sigma_{z}|\widetilde{\beta}\rangle\frac{\omega_{l}}{2\pi}\int_{0}^{2\pi/\omega_{l}}\frac{1}{2}C_{j}^{z}(t)e^{-in\omega_{l}t}dt\nonumber \\
 &  & +\langle\widetilde{\alpha}|\sigma_{-}|\widetilde{\beta}\rangle\frac{\omega_{l}}{2\pi}\int_{0}^{2\pi/\omega_{l}}C_{j}^{+}(t)e^{-in\omega_{l}t}dt\nonumber \\
 &  & +\langle\widetilde{\alpha}|\sigma_{+}|\widetilde{\beta}\rangle\frac{\omega_{l}}{2\pi}\int_{0}^{2\pi/\omega_{l}}C_{j}^{-}(t)e^{-in\omega_{l}t}dt.\label{eq:A4}
\end{eqnarray}
where $\langle\widetilde{\alpha}|\sigma_{z}|\widetilde{\beta}\rangle\equiv c_{\alpha\beta}$ and $\langle\widetilde{\alpha}|\sigma_{-}|\widetilde{\beta}\rangle\equiv d_{\alpha\beta}$ can be easily derived.
Therefore, the remaining task is to determine the explicit form for $C^{\lambda}_j(t)$ according to the operator $\hat{X}^{(j)}$ and integrals in Eq.~(\ref{eq:A4}).

When $\hat{X}^{(j)}=\sigma_z/2$, $C_j^{\lambda}(t)$ can be determined as follows:
\begin{eqnarray}
C_{z}^{z}(t) & = & \frac{1}{2}\mathrm{Tr}[\sigma_{z}e^{S(t)}\sigma_{z}e^{-S(t)}]=\cos\left[\frac{A\xi}{\omega_{l}}\sin(\omega_{l}t)\right],\\
C_{z}^{+}(t) & = & \frac{1}{2}\mathrm{Tr}[R^{\dagger}(t)\sigma_{+}R(t)e^{S(t)}\sigma_{z}e^{-S(t)}]\nonumber \\
 & = & \frac{1}{2}\mathrm{Tr}[e^{-i\omega_{l}t}\sigma_{+}e^{S(t)}\sigma_{z}e^{-S(t)}]=\frac{1}{2}ie^{-i\omega_{l}t}\sin\left[\frac{A\xi}{\omega_{l}}\sin(\omega_{l}t)\right],\\
C_{z}^{-}(t) & = & \frac{1}{2}\mathrm{Tr}[e^{i\omega_{l}t}\sigma_{-}e^{S(t)}\sigma_{z}e^{-S(t)}]=-\frac{1}{2}ie^{i\omega_{l}t}\sin\left[\frac{A\xi}{\omega_{l}}\sin(\omega_{l}t)\right].
\end{eqnarray}
By using the following identities:
\begin{eqnarray}
\frac{\omega_{l}}{2\pi}\int_{0}^{2\pi/\omega_{l}}\cos\left[\frac{A\xi}{\omega_{l}}\sin(\omega_{l}t)\right]e^{-in\omega_{l}t}dt & = & J_{0}\left(\frac{A\xi}{\omega_{l}}\right)\delta_{n,0}\nonumber \\
 &  & +\sum_{k=1}^{\infty}J_{2k}\left(\frac{A\xi}{\omega_{l}}\right)(\delta_{n,2k}+\delta_{n,-2k}),\\
\frac{\omega_{l}}{2\pi}\int_{0}^{2\pi/\omega_{l}}i\sin\left[\frac{A\xi}{\omega_{l}}\sin(\omega_{l}t)\right]e^{-in\omega_{l}t}dt & = & \sum_{k=1}^{\infty}J_{2k-1}\left(\frac{A\xi}{\omega_{l}}\right)(\delta_{n,2k-1}-\delta_{n,-2k+1}),
\end{eqnarray}
we obtain the explicit form for $X^z_{\alpha\beta,n}$.

\bibliography{mybib}

\begin{figure}
  \includegraphics[width=12cm]{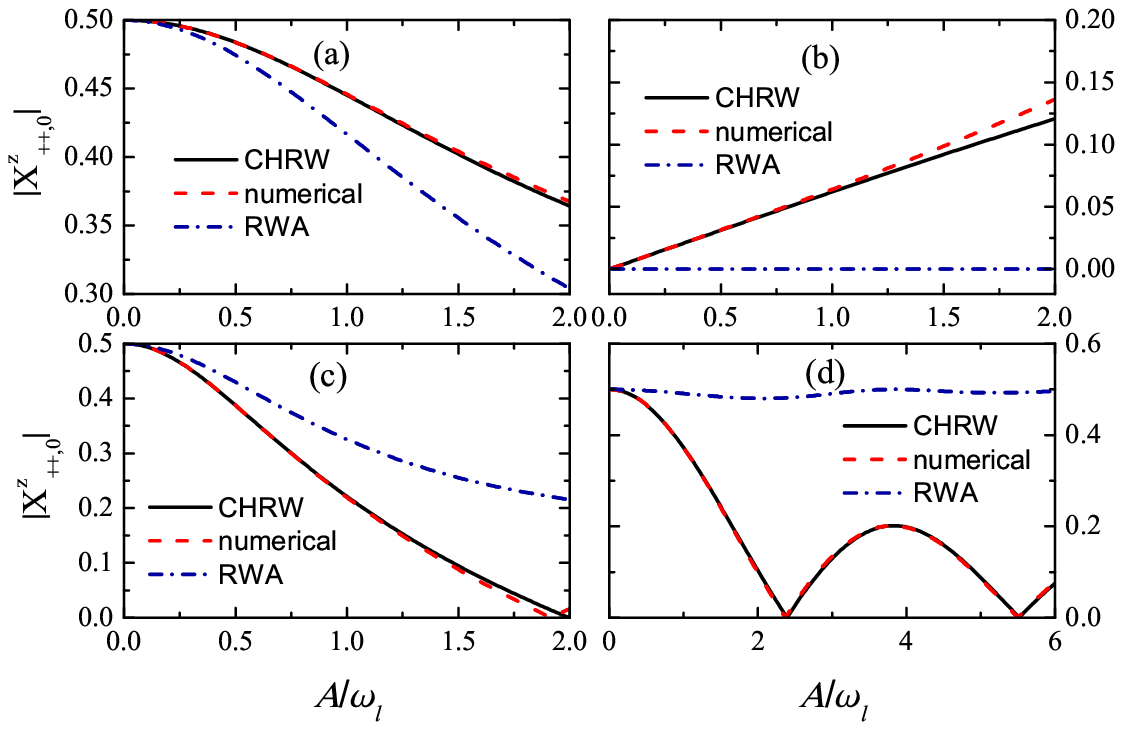}\\
  \caption{(Color online) The coefficient $|X^z_{++,0}|$ as a function of the ratio $A/\omega_l$ for various driving frequencies: (a) $\omega_l=0.5\omega_0$, (b) $\omega_l=\omega_0$, (c) $\omega_l=1.5\omega_0$, and (d) $\omega_l=5\omega_0$.}\label{fig2}
\end{figure}

\begin{figure}
  \includegraphics[width=12cm]{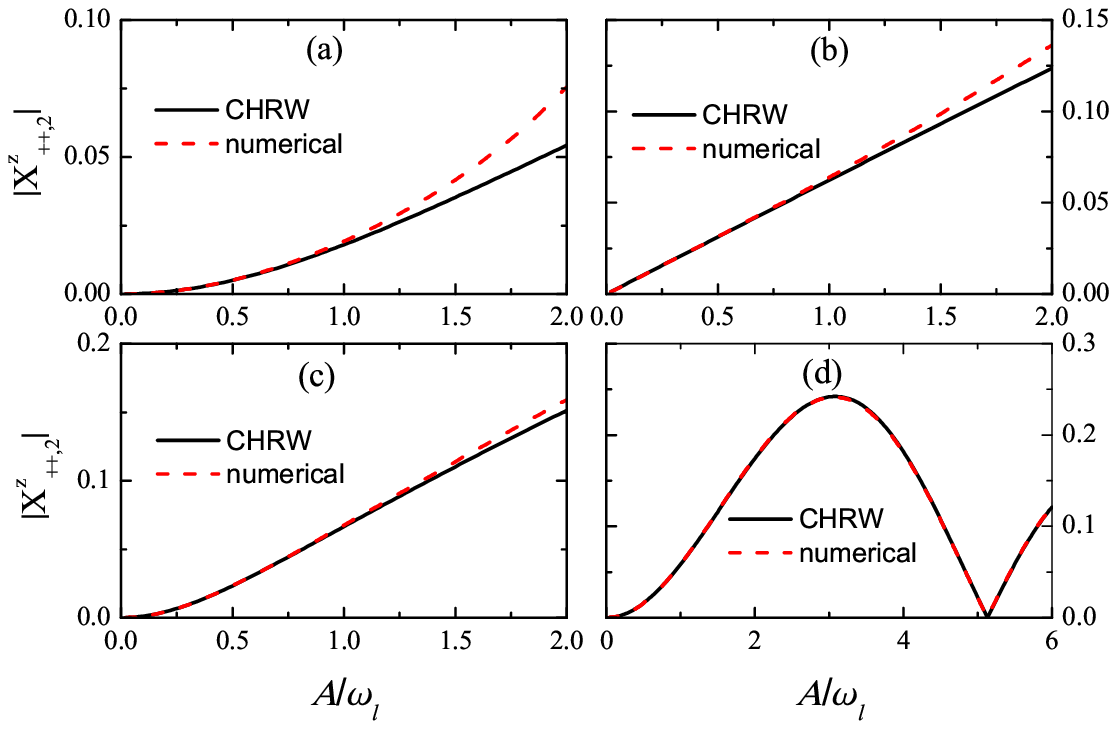}\\
  \caption{(Color online) The coefficient $|X^z_{++,2}|$ as a function of the ratio $A/\omega_l$ for various driving frequencies: (a) $\omega_l=0.5\omega_0$, (b) $\omega_l=\omega_0$, (c) $\omega_l=1.5\omega_0$, and (d) $\omega_l=5\omega_0$.}\label{fig3}
\end{figure}

\begin{figure}
  \includegraphics[width=8cm]{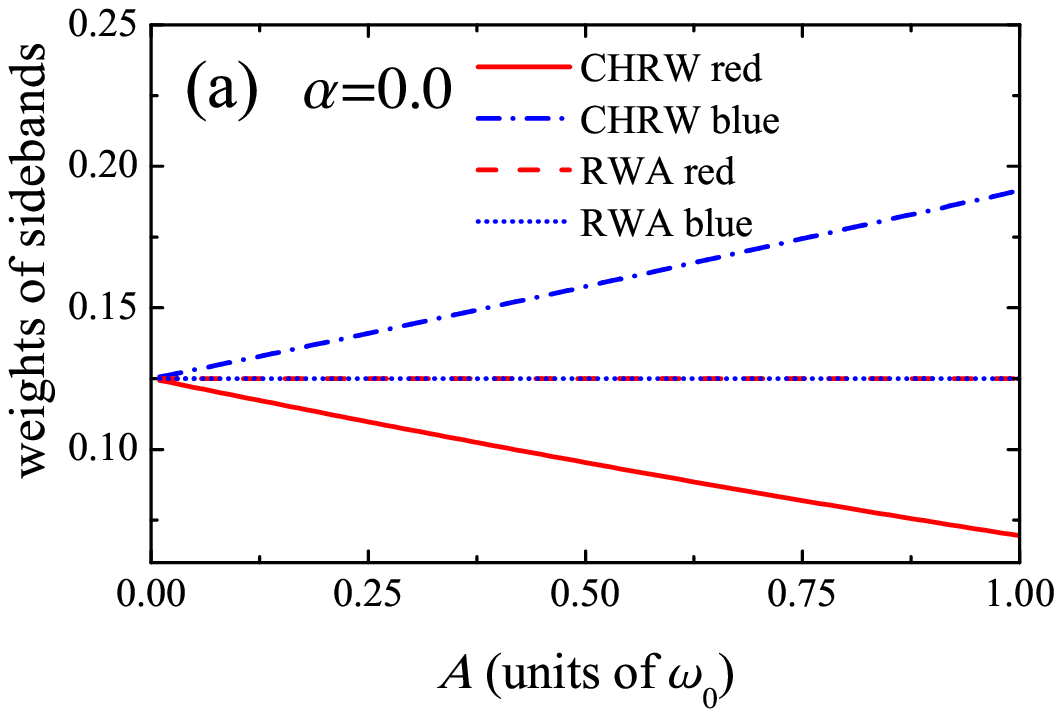}
  \includegraphics[width=8cm]{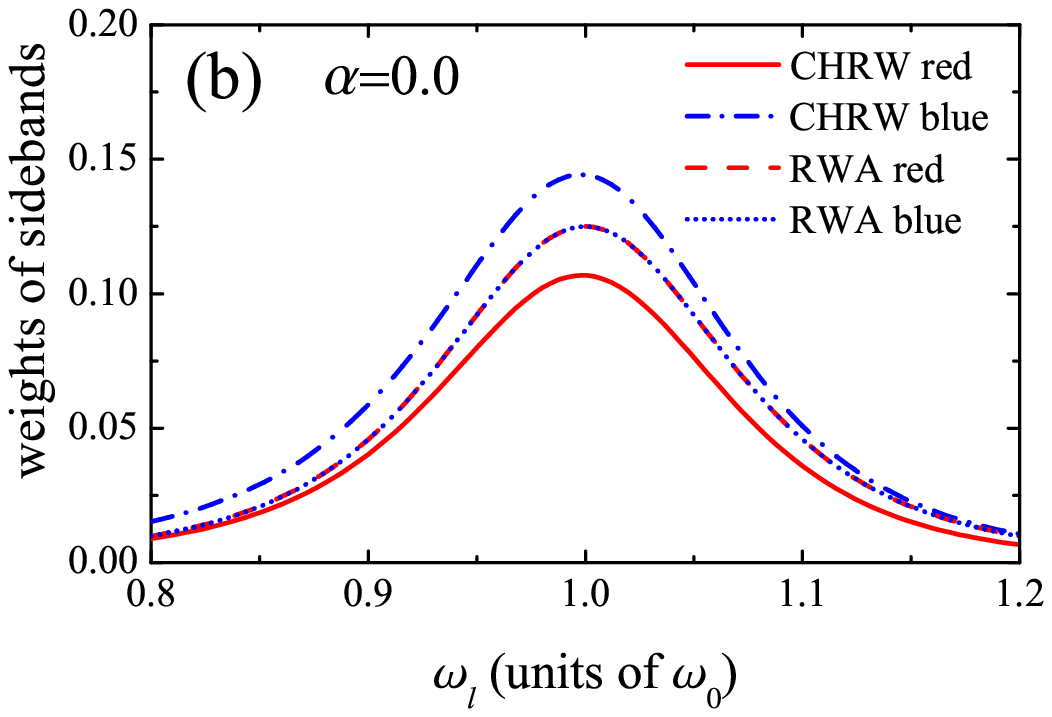}\\
  \caption{(Color online) (a) The weights of the sidebands in the first-order triplet as a function of $A$ for $\omega_l=\omega_0$. (b) The weights of the sidebands in the first-order triplet as a function of $\omega_l$ for $A=0.3\omega_0$. The radiative decay rate is $\kappa=0.02\omega_0$.}\label{fig4}
\end{figure}

\begin{figure}
  \includegraphics[width=8cm]{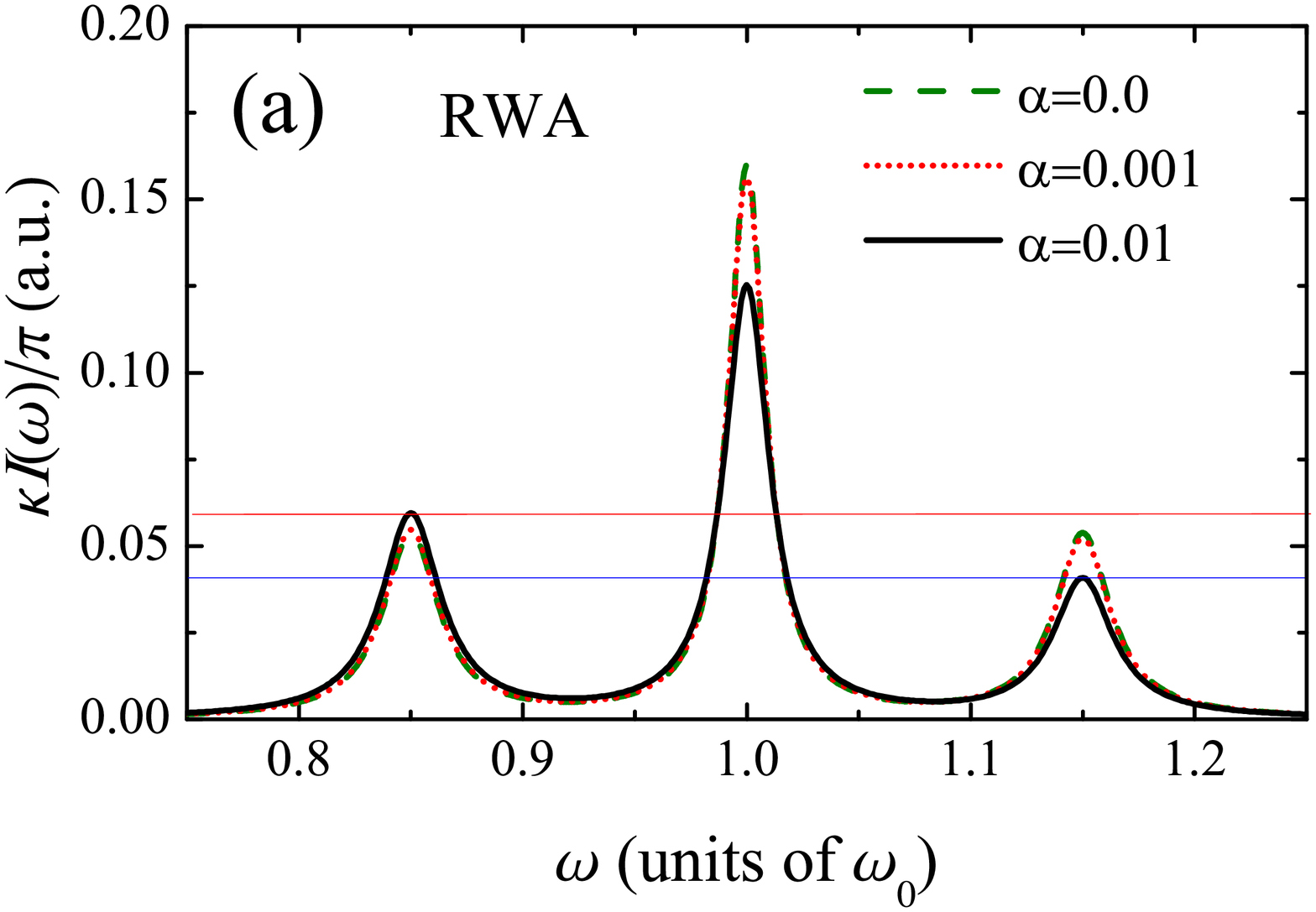}
  \includegraphics[width=8cm]{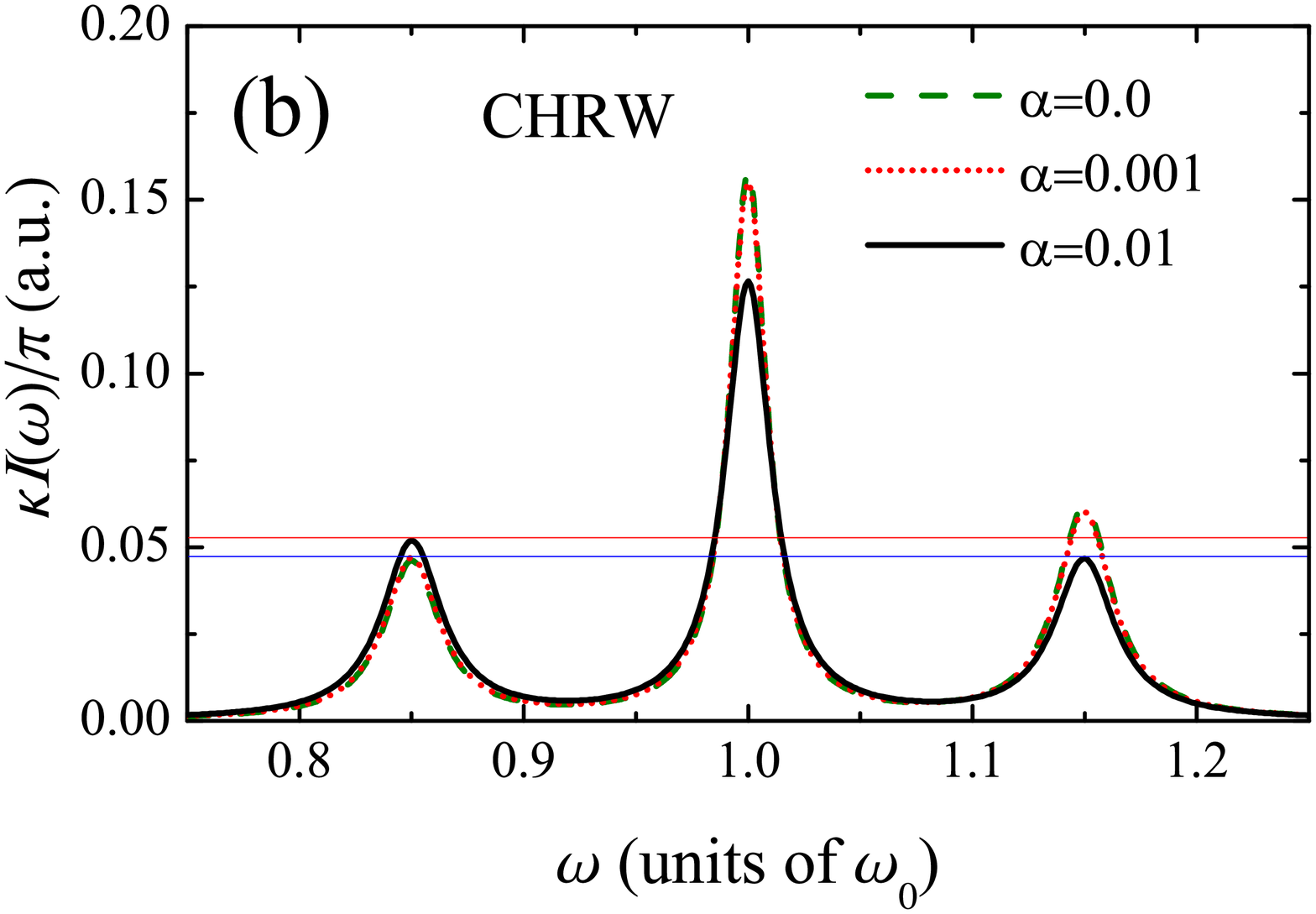}\\
  \includegraphics[width=8cm]{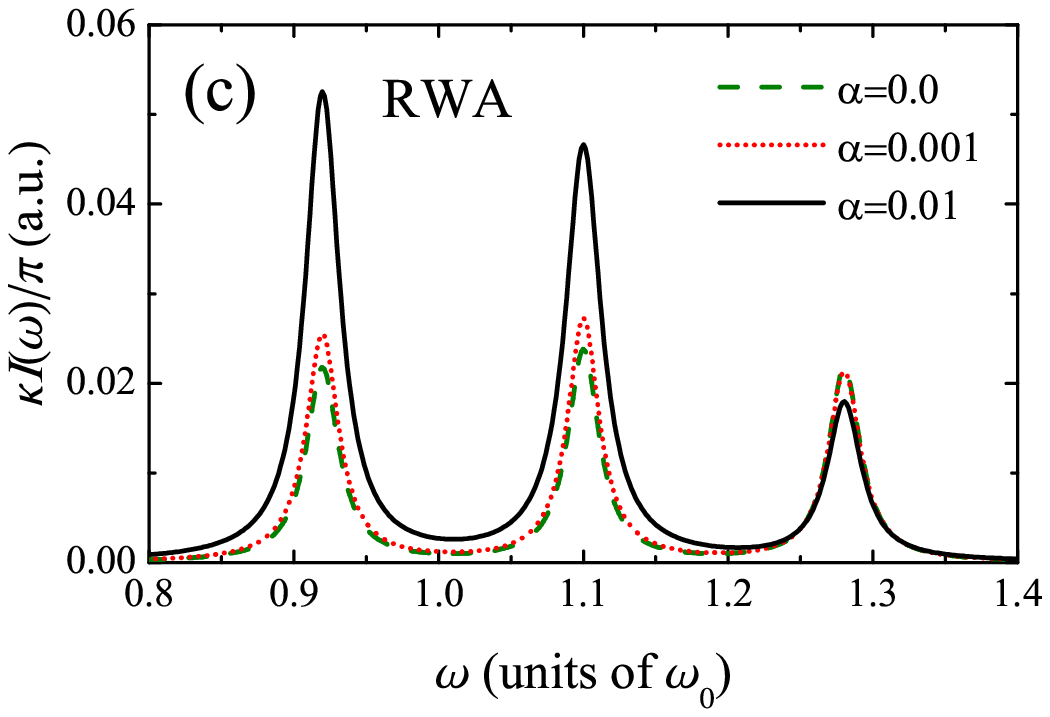}
  \includegraphics[width=8cm]{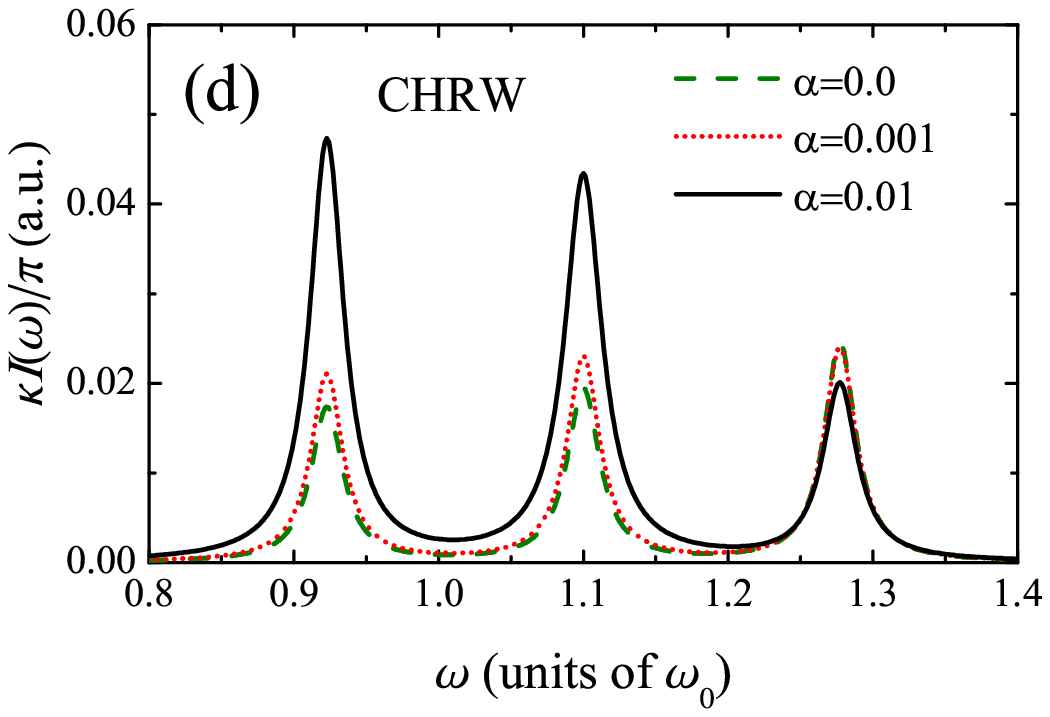}\\
  \caption{(Color online) The spectrum $I(\omega)$ of the resonance and nonresonance fluorescence for $A=0.3\omega_0$ and $\kappa=0.02\omega_0$. The driving frequencies are set as $\omega_l=\omega_0$ for (a) and (b); $\omega_l=1.1\omega_0$ for (c) and (d).}\label{fig5}
\end{figure}

\begin{figure}
  \includegraphics[width=8cm]{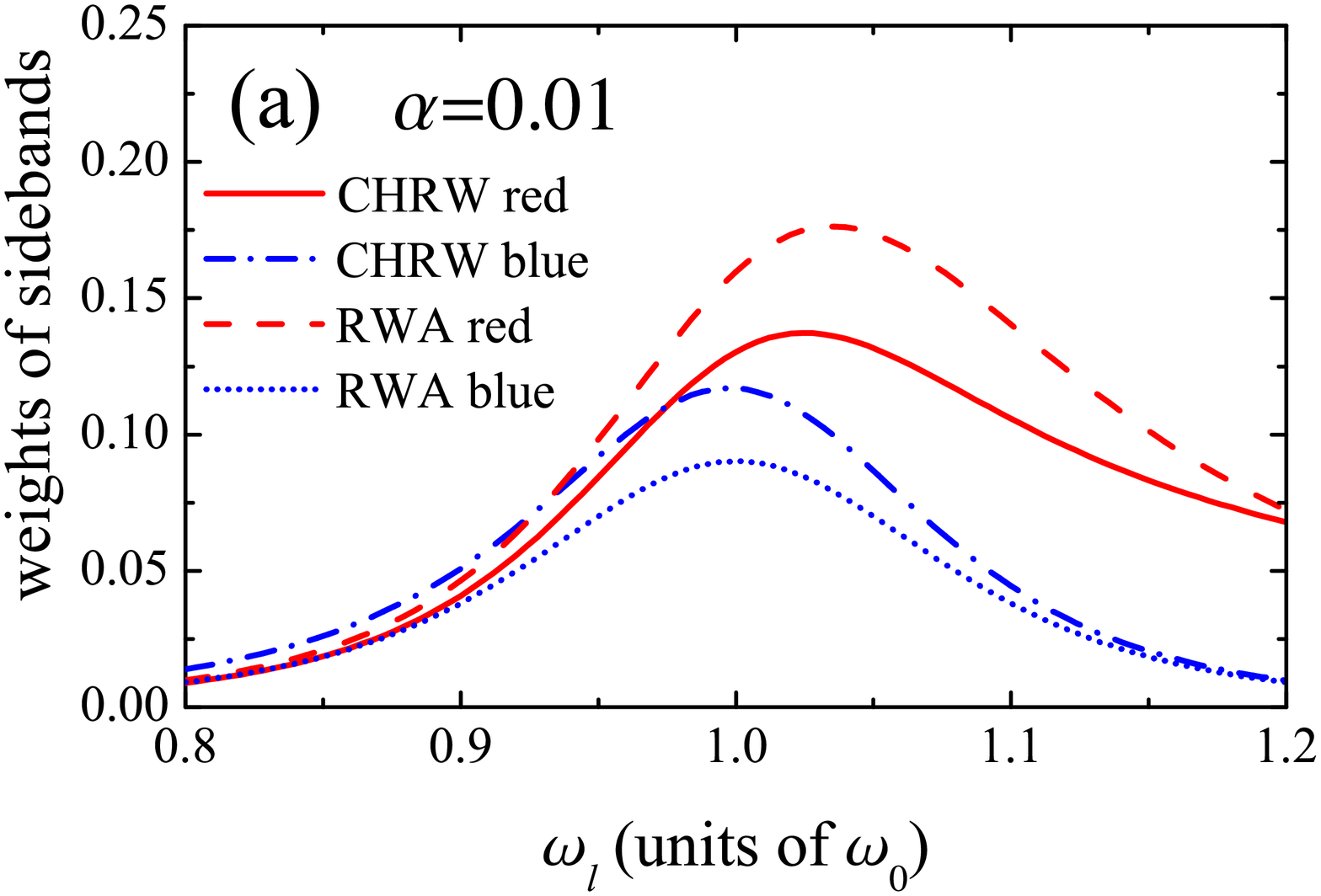}
  \includegraphics[width=8cm]{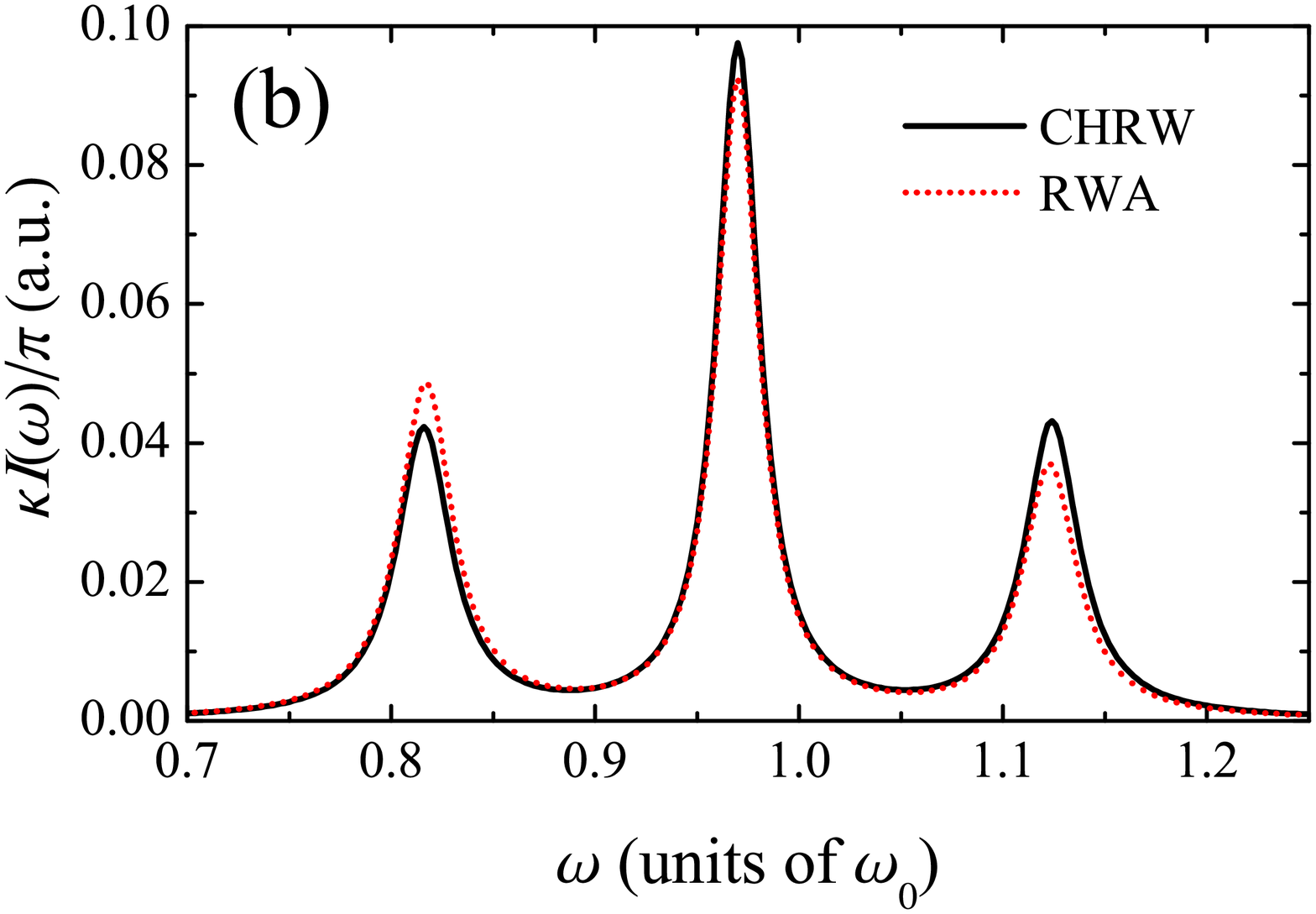}\\
  \caption{(Color online) (a) The weights of the sidebands in the first-order triplet as a function of $\omega_l$ for $A=0.3\omega_0$ and $\kappa=0.02\omega_0$. (b) The spectrum $I(\omega)$ for $\omega_l=0.97\omega_0$. The other parameters are the same as (a).}\label{fig6}
\end{figure}

\begin{figure}
  \includegraphics[width=8cm]{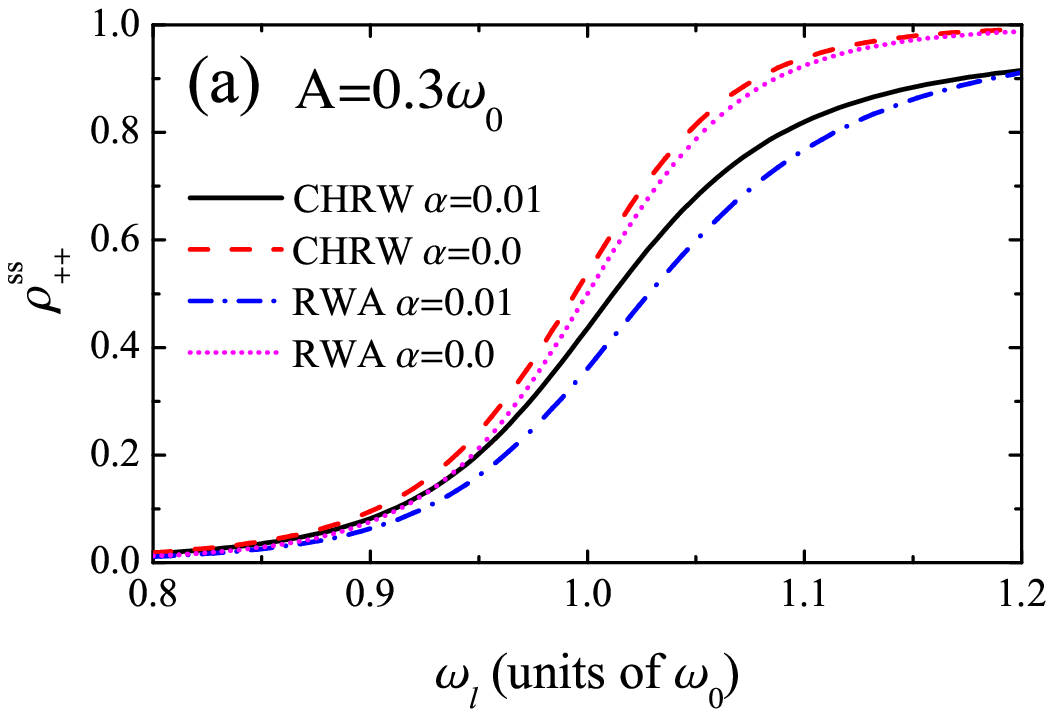}
  \includegraphics[width=8cm]{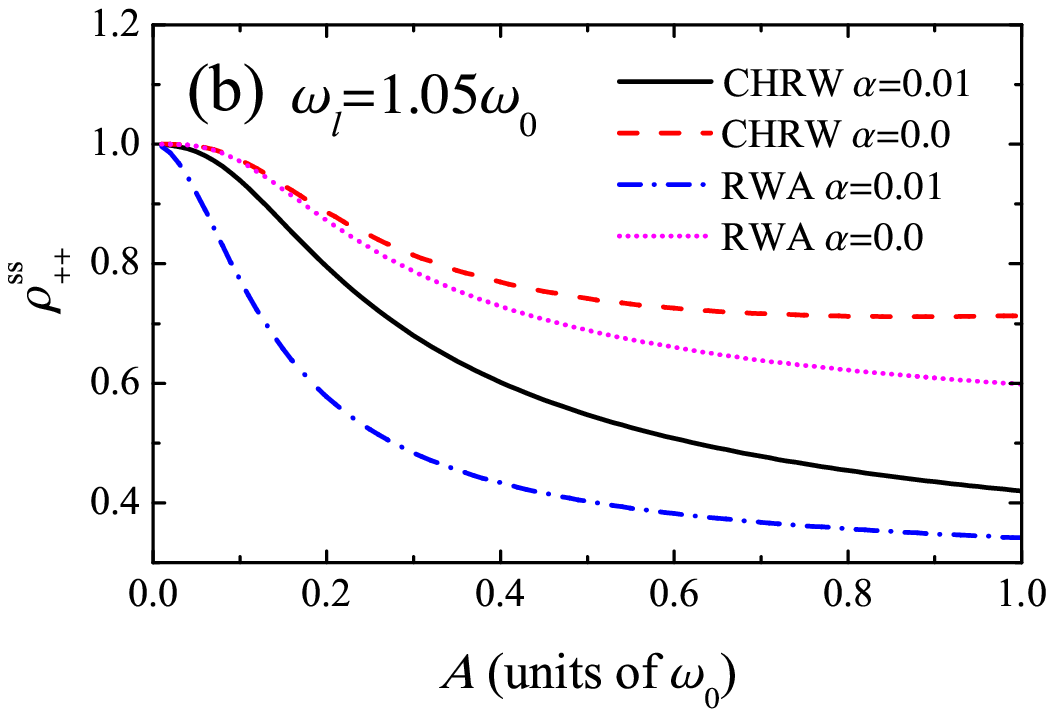}\\
  \caption{(Color online) (a) The steady Floquet-state population $\rho_{++}^{\mathrm{ss}}$ as a function of $\omega_l$. (b) The steady Floquet-state population $\rho_{++}^{\mathrm{ss}}$ as a function of $A$. The radiative decay rate is $\kappa=0.02\omega_0$.}\label{fig7}
\end{figure}

\begin{figure}
  \includegraphics[width=8cm]{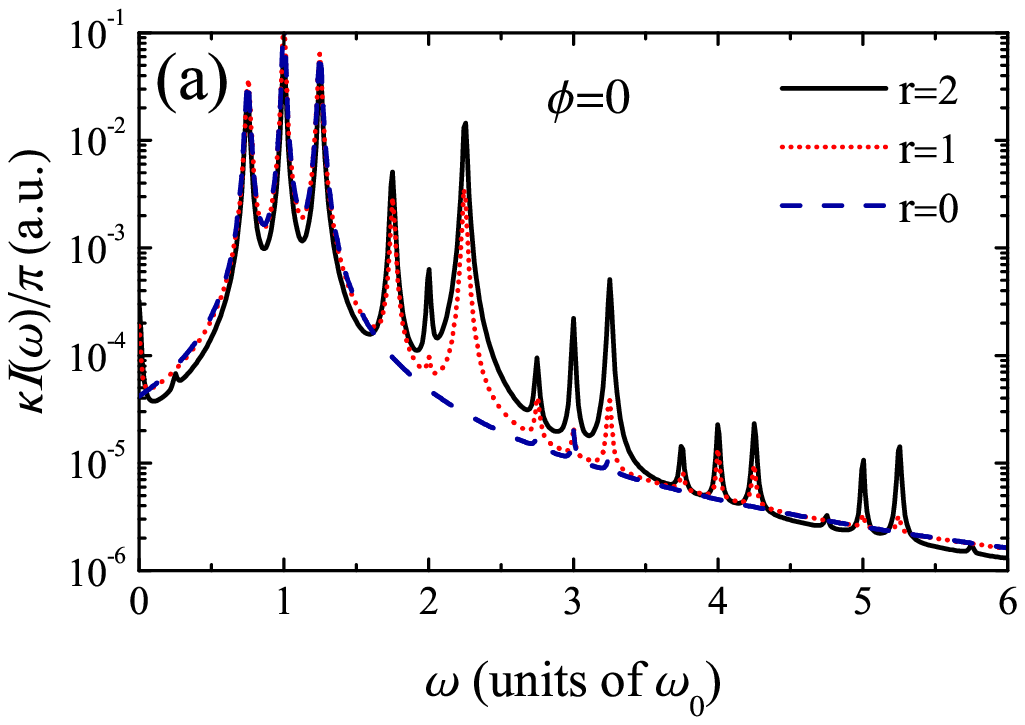}
  \includegraphics[width=8cm]{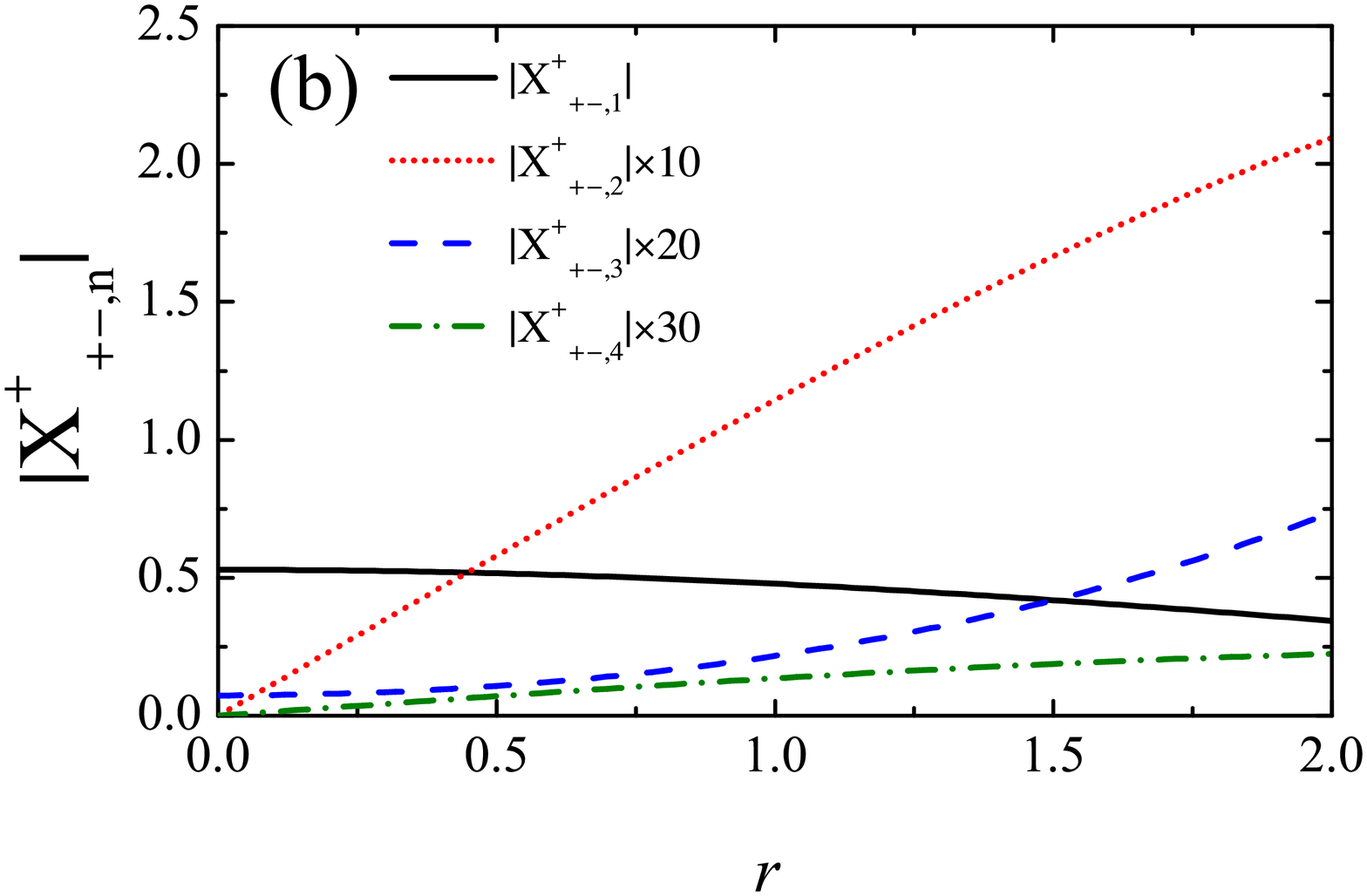}\\
  \caption{(Color online) (a) The fluorescence spectrum $I(\omega)$ of the biharmonically driven TLS for $\omega_l=\omega_0$, $A=0.5\omega_0$, $\kappa=0.02\omega_0$, and $\alpha=0$. (b) $|X^+_{+-,n}|$ for $n=1,2,3,4$ as a function of relative phase $r$. The other parameters are the same as (a).}\label{fig8}
\end{figure}

\begin{figure}
  \includegraphics[width=8cm]{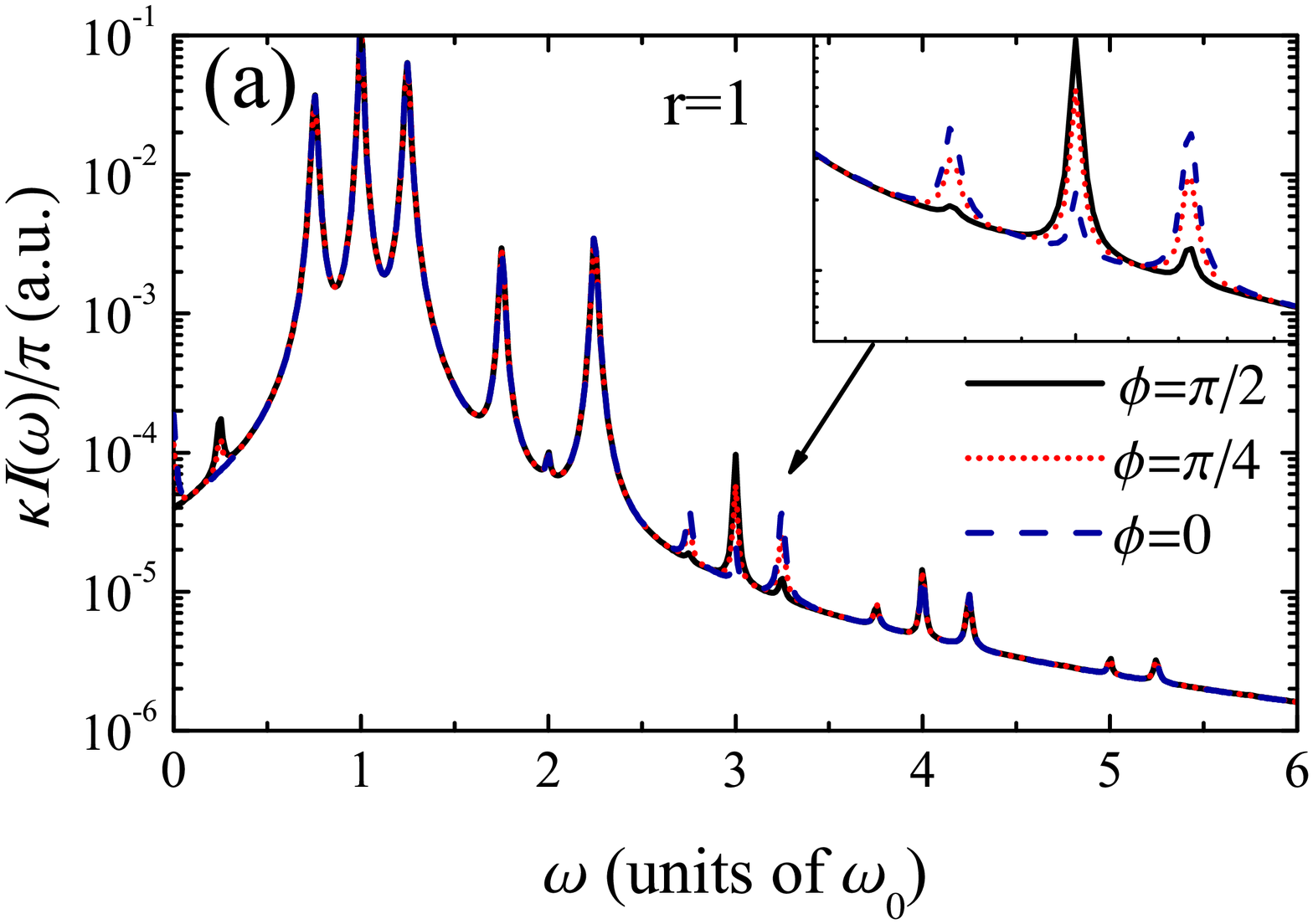}
  \includegraphics[width=8cm]{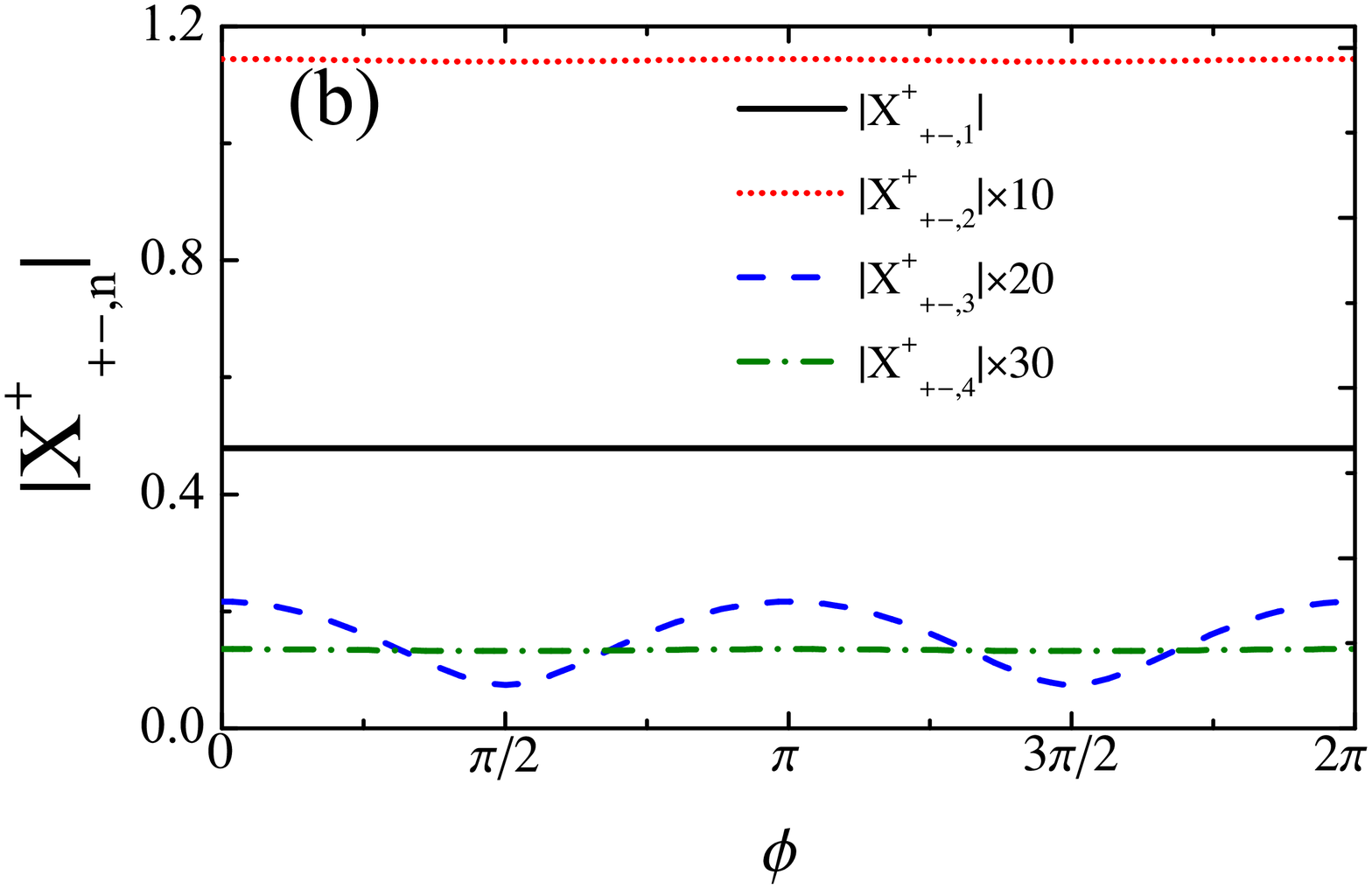}
  \caption{(Color online) (a) The fluorescence spectrum $I(\omega)$ of the biharmonically driven TLS for $\omega_l=\omega_0$, $A=0.5\omega_0$, $\kappa=0.02\omega_0$, and $\alpha=0$. (b) $|X^+_{+-,n}|$ for $n=1,2,3,4$ as a function of relative phase $\phi$ for $r=1$. The other parameters are the same as (a).}\label{fig9}
\end{figure}

\begin{figure}
  \includegraphics[width=8cm]{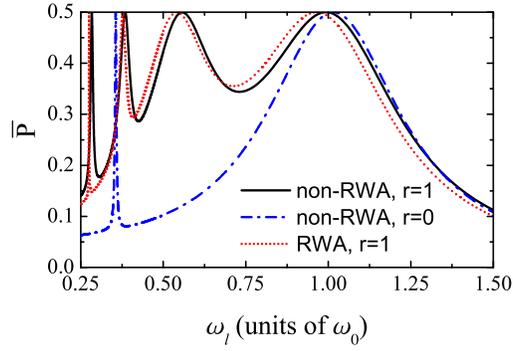}
  \caption{(Color online) The time-averaged transition probability of the biharmonically driven TLS as a function of driving frequency $\omega_l$ for $A=0.5\omega_0$ and $\phi=0$. The case of $r=0$ corresponds to the harmonic driving.}\label{fig10}
\end{figure}

\begin{figure}
  \includegraphics[width=8cm]{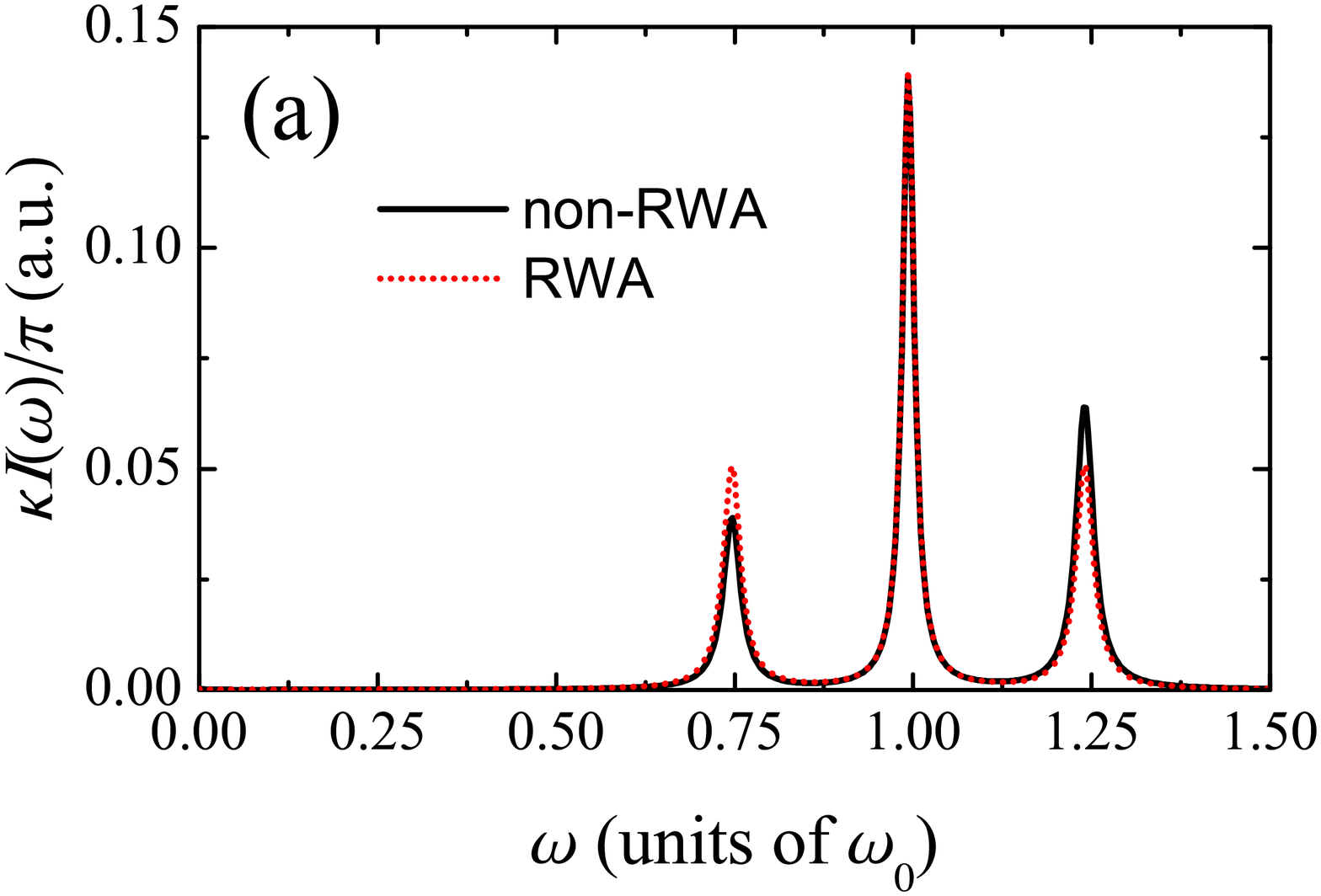}
  \includegraphics[width=8cm]{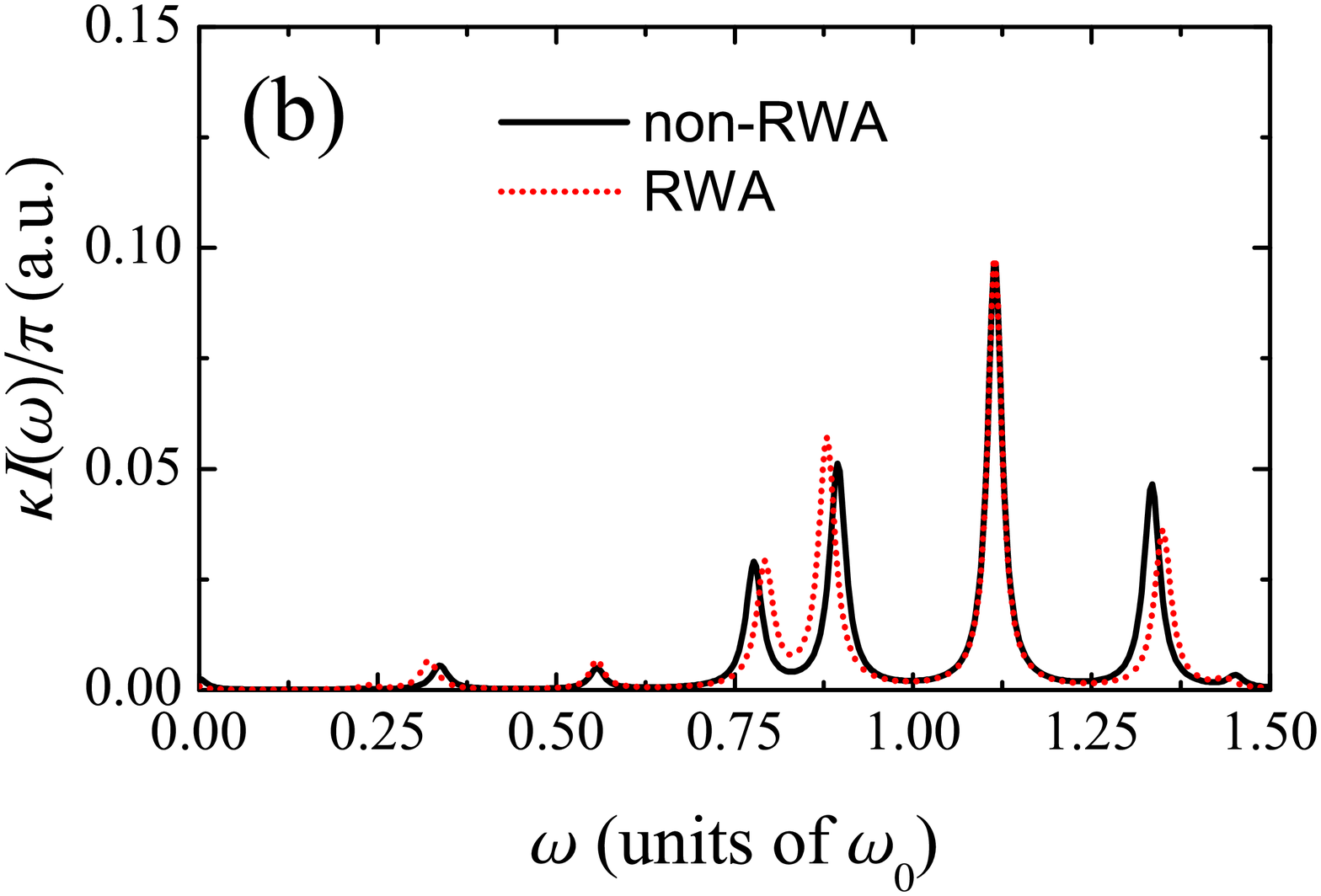}\\
  \includegraphics[width=8cm]{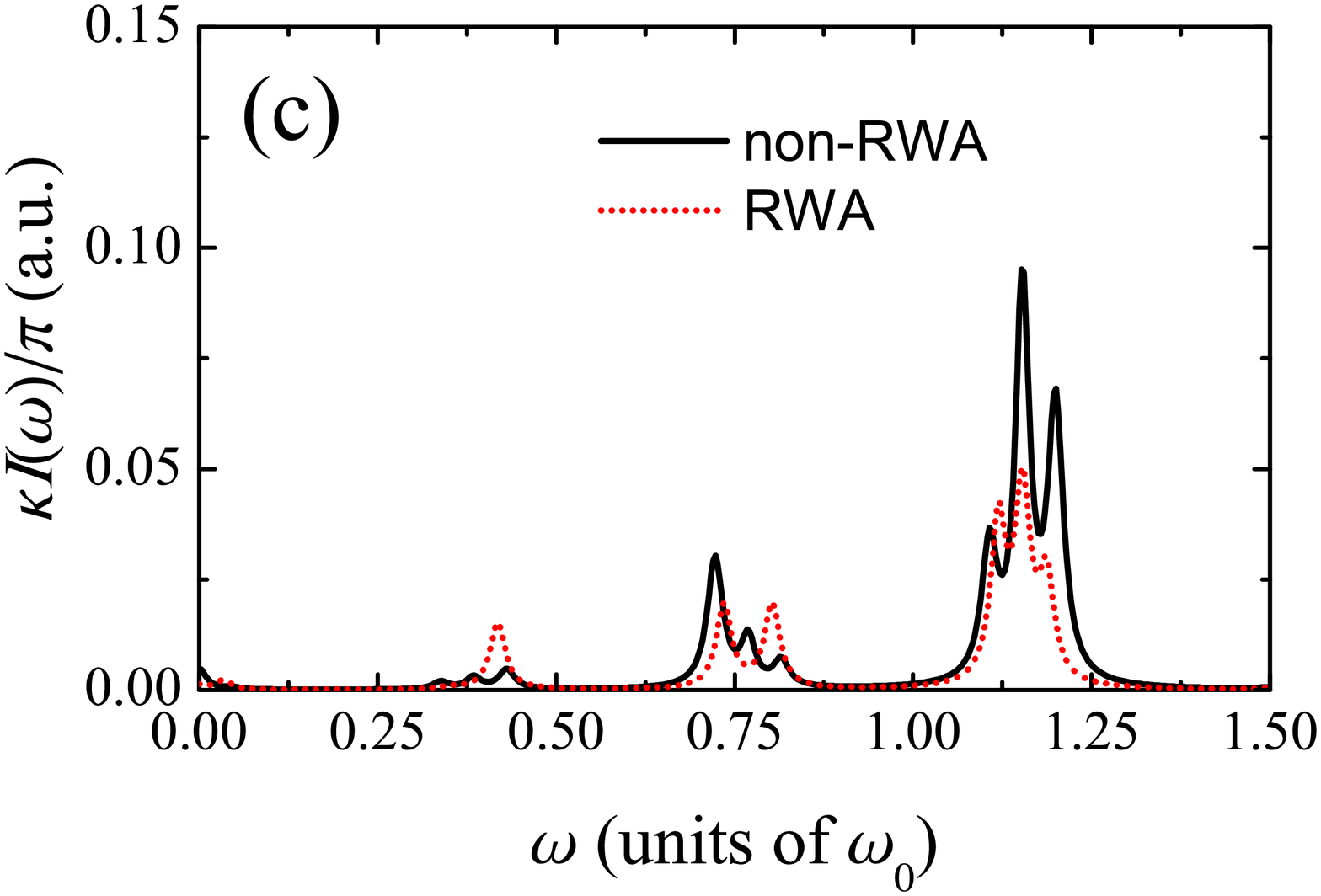}
  \includegraphics[width=8cm]{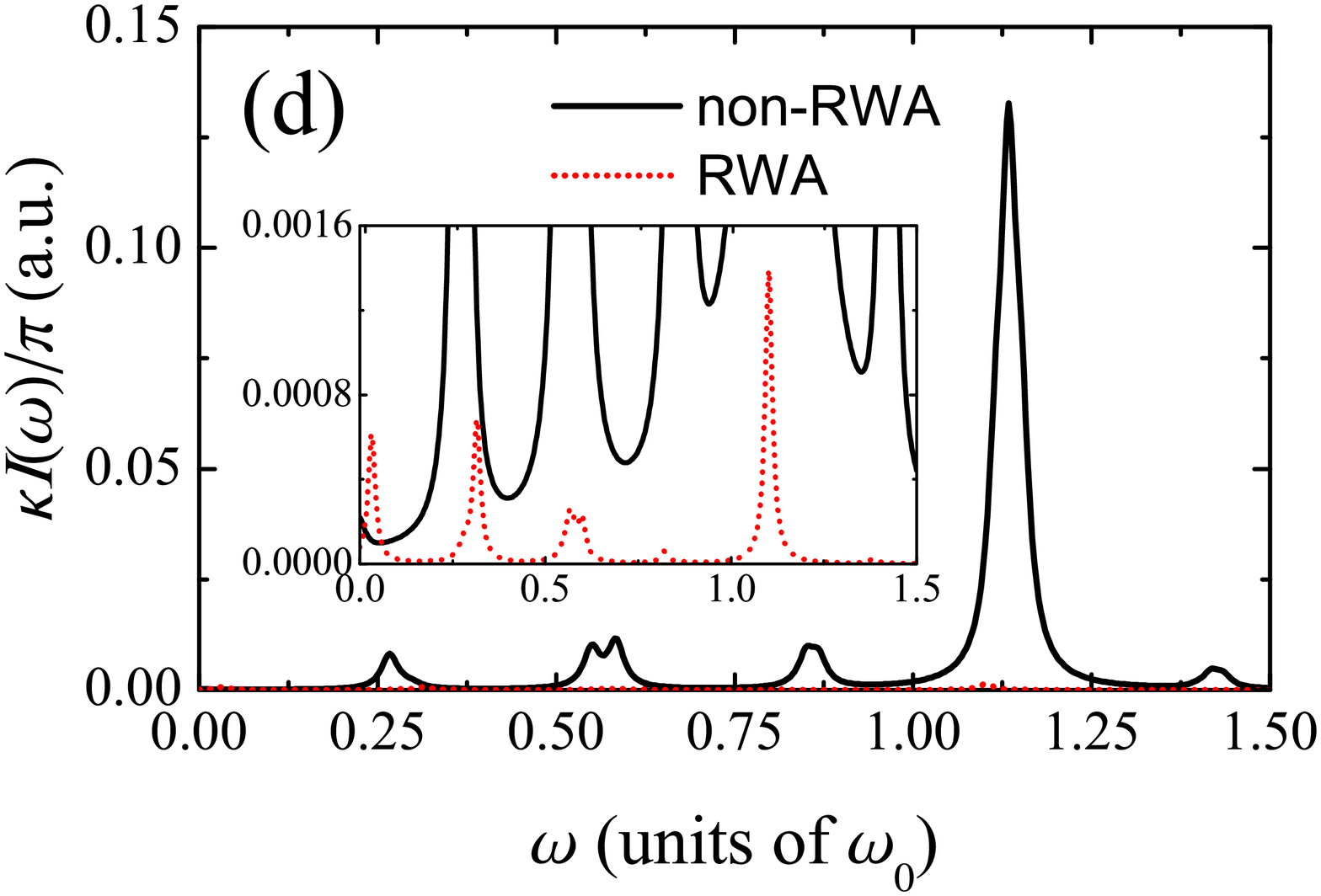}\\
  \caption{(Color online) The resonance fluorescence spectrum as a function of $\omega$ under various resonance conditions of the biharmonically driven TLS for $\kappa=0.02\omega_0$ and $\alpha=0$. The resonance frequencies are given by (a) $\omega_l=0.9933\omega_0$, (b) $\omega_l=0.5572\omega_0$, (c) $\omega_l=0.3844$, and (d) $\omega_l=0.2834$ for $A=0.5\omega_0$, $r=1$, and $\phi=0$.}\label{fig11}
\end{figure}

\begin{figure}
  \includegraphics[width=8cm]{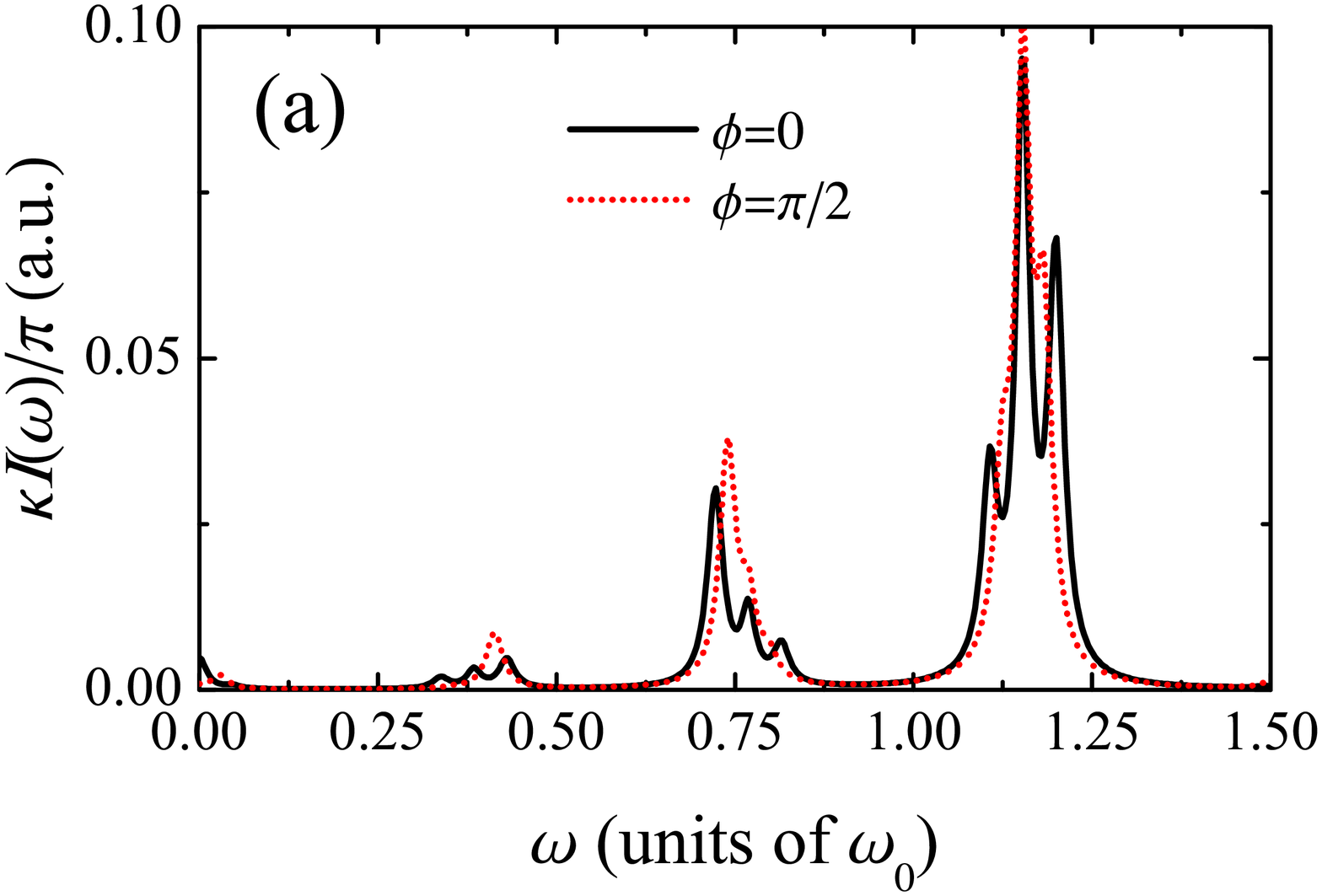}
  \includegraphics[width=8cm]{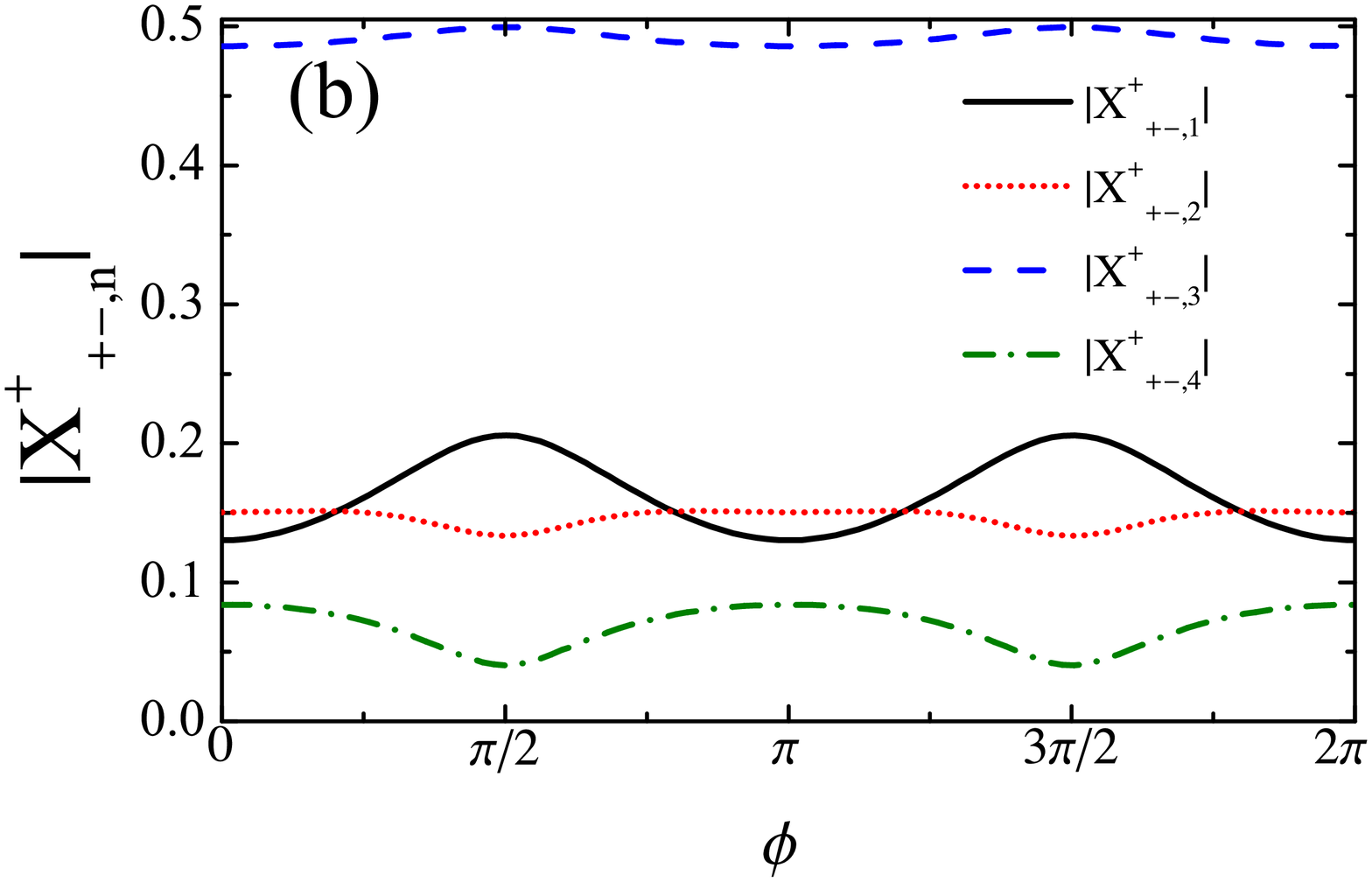}
  \caption{(Color online) (a) The resonance fluorescence spectrum as a function of $\omega$ for $\phi=0$ and $\phi=\pi/2$. The other parameters are given as Fig.~\ref{fig11}(c). (b) $X^+_{+-,n}$ with $n=1,2,3,4$ as a function of $\phi$ for $r=1$. The other parameters are the same as (a).}\label{fig12}
\end{figure}

\begin{figure}
  \includegraphics[width=8cm]{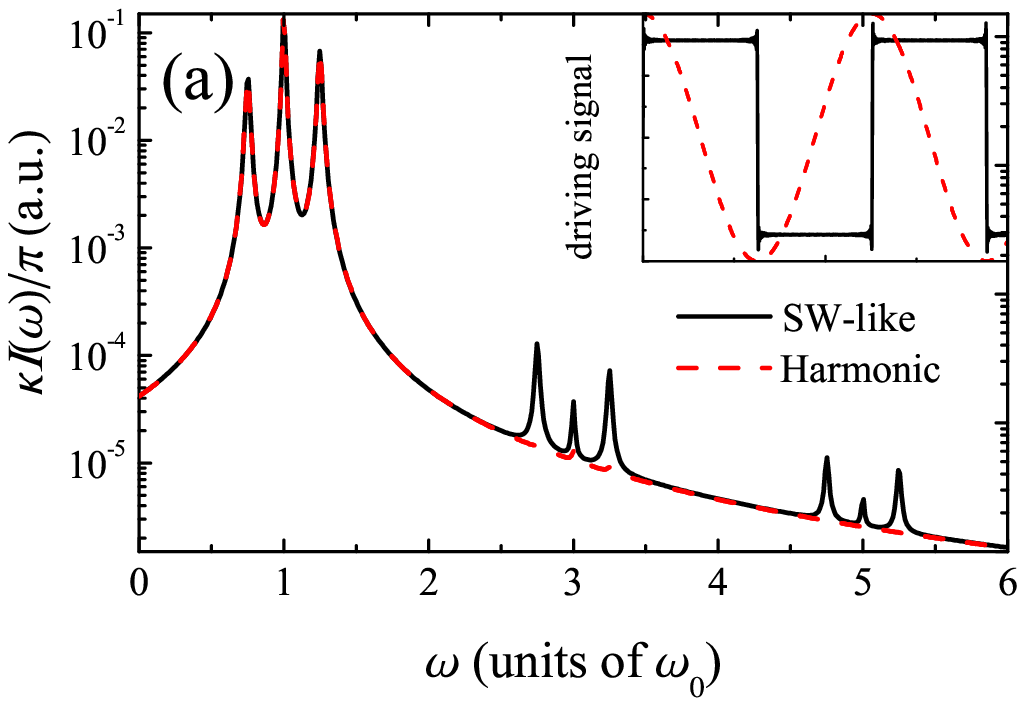}
  \includegraphics[width=8cm]{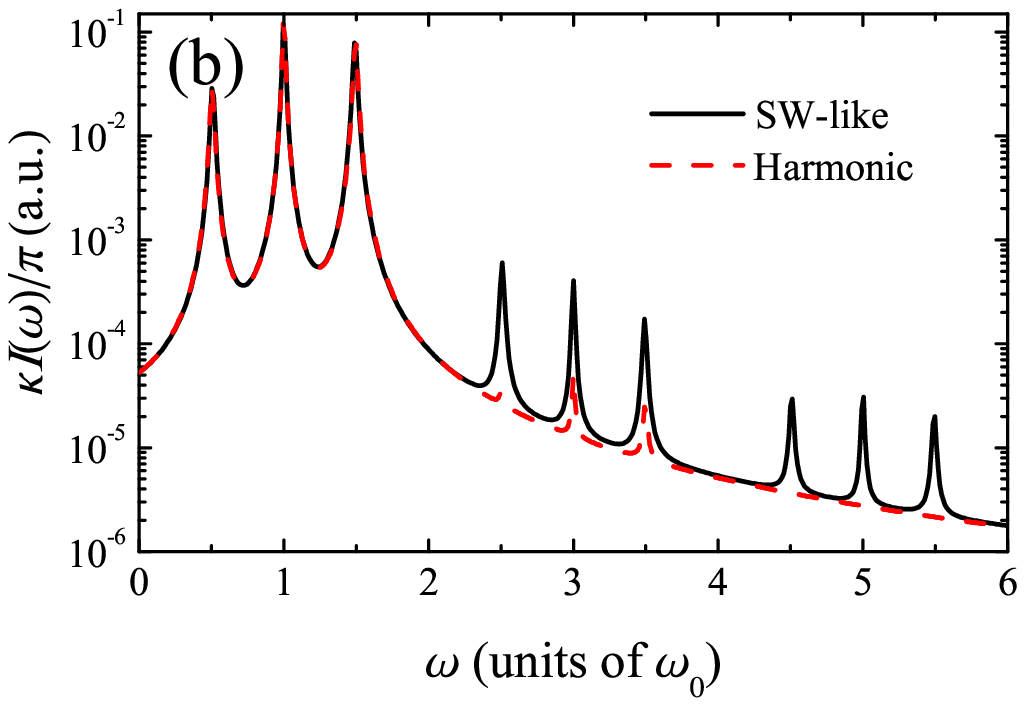}\\
  \caption{(Color online) The fluorescence spectrum $I(\omega)$ of the multiharmonically driven TLS for $\omega_l=\omega_0$, $\kappa=0.02\omega_0$, and $\alpha=0$. The driving strength is set: (a) $A=0.5\omega_0$ and (b) $A=\omega_0$. The inset of left panel is the comparison of driving signals.}\label{fig13}
\end{figure}

\end{document}